\documentclass[journal]{IEEEtran}
\usepackage{latexsym}
\usepackage{algorithm,algorithmic}
\usepackage{array}
\usepackage{mdwmath}
\makeatletter
\newcommand*{\rom}[1]{\expandafter\@slowromancap\romannumeral #1@}
\usepackage{mdwtab}
\usepackage{eqparbox}
\usepackage{stfloats}
\usepackage{latexsym}
\usepackage{amssymb}
\usepackage{amsmath}
\usepackage{graphicx}
\usepackage{csquotes}
\usepackage{balance}
\usepackage[justification=centering]{caption}
\DeclareGraphicsExtensions{.eps,.pdf,.gif,.png,.jpg}
\usepackage{color}
\usepackage[breaklinks=true]{hyperref}
\usepackage{breakcites}
\usepackage{url}  
\usepackage{amsthm}
\usepackage{multirow}
\usepackage{color}
\usepackage{epstopdf}
\usepackage{endnotes}
\usepackage{framed}
\usepackage{cool} % for partial derivative
\usepackage{enumerate} % for \begin{enumerate}[(a)], or [a)]
\usepackage{url}
\usepackage[subnum]{cases}
\usepackage[nocomma]{optidef}
\usepackage{etoolbox}
\usepackage{cite} 
\usepackage{caption}
\usepackage{subcaption}

\ifCLASSINFOpdf
\else
\fi
\begin{document}
\title{A Business Model for Resource Sharing in Cell-Free UAVs-Assisted Wireless Networks}

\author{Yan~Kyaw~Tun,~
   	    Yu~Min~Park,
   	    Tra~Huong~Thi~Le, 
	    Zhu~ Han,~\IEEEmembership{Fellow,~IEEE,~}\\
	and~Choong~Seon~Hong,~\IEEEmembership{Senior~Member,~IEEE}

\thanks{Yan Kyaw Tun, Yu Min Park, Tra Huong Thi Le and Choong Seon Hong  are with the Department of Computer Science and Engineering, Kyung Hee University,  Yongin-si, Gyeonggi-do 17104, Rep. of Korea, e-mail:{\{ykyawtun7, yumin096, huongtra25, cshong\}@khu.ac.kr}.}
%\thanks{Nguyen H. Tran is with the School of Information Technologies, The University of Sydney, NSW 2006, Australia, email{\{nguyen.tran\}@sydney.edu.au}.}
%\thanks{Duy Trong Ngo is with the School of Electrical Engineering and Computing, The University of Newcastle, Callaghan, NSW 2308, Australia, email{\{duy.ngo\}@newcastle.edu.au}.}
\thanks{Zhu Han is with the Electrical and Computer Engineering Department,
    	University of Houston, Houston, TX 77004, and the Department of
    	Computer Science and Engineering, Kyung Hee University, Yongin-si,
    	Gyeonggi-do 17104,  Rep. of Korea, email{\{zhan2\}@uh.edu}.}}

\maketitle
%As a general rule, do not put math, special symbols or citations
\begin{abstract}
Unmanned aerial vehicles (UAVs) are widely deployed to enhance the wireless network capacity and to provide communication services to mobile users beyond the infrastructure coverage. Recently, with the help of a promising technology called network virtualization, multiple service providers (SPs) can share the infrastructures and wireless resources owned by the mobile network operators (MNOs). Then, they provide specific services to their mobile users using the resources obtained from MNOs. However, wireless resource sharing among SPs is challenging as each SP wants to maximize their utility/profit selfishly while satisfying the QoS requirement of their mobile users. Therefore, in this paper, we propose joint user association and wireless resource sharing problem in the cell-free UAVs-assisted wireless networks with the objective of maximizing the total network utility of the SPs while ensuring QoS constraints of their mobile users and the resource constraints of the UAVs deployed by MNOs. To solve the proposed mixed-integer non-convex problem, we decompose the proposed problem into two subproblems: users association, and resource sharing problems. Then, a two-sided matching algorithm is deployed in order to solve users association problem. We further deploy the whale optimization and Lagrangian relaxation algorithms to solve the resource sharing problem. Finally, extensive numerical results are provided in order to show the effectiveness of our proposed algorithm.
\end{abstract}


% \fi
\IEEEpeerreviewmaketitle

\section{Introduction}
\IEEEPARstart{W}{ith} the help of the wireless network virtualization technology, multiple services providers (SPs) can provide specific services (e.g., messaging, video streaming, online gaming, and so on) to their mobile users by sharing wireless communication infrastructures (e.g., terrestrial base stations (BSs) including macro BSs and small-cell BSs, access points, cell sites, etc.,) and wireless resources, such as bandwidth and power, owned by the mobile network operators (MNOs) \cite{liang2014wireless}\cite{haider2009challenges}. As a result, SPs can reduce their capital expenditure (CAPEX) and operational expenditure (OPEX). However, it is challenging to efficiently share wireless resources among SPs without interfering each others because of selfishness of SPs. Therefore, in \cite{tun2019wireless}, authors have proposed an efficient two-level resource sharing framework in the wireless resource virtualization. In the upper-level, an MNO efficiently share its wireless resources, i.e., bandwidth and power to multiple SPs while ensuring inter-isolation among SPs, and then, in the lower-level, SPs allocates the resource it received from the MNO to its mobile users efficiently. Finally, authors deployed generalized Kelly mechanism and Karush-Kuhn-Tucker (KKT) conditions in order to address both upper-level and lower-level problems.

According to \cite{linsa}, only 63.2\% of world's population can get internet access till Oct 2020, so \emph{ the remaining 36.8\% are out of the internet coverage}. Consequently, researchers from both academic and industry are avid to deploy unmanned aerial vehicles (UAVs) such as drones, balloons, and airships as communication and computation platforms in order to bring one-third of the world's population back into the Internet coverage and increase the global connectivity. Due to the flexibility and cost effective deployment, UAVs are also disposed in the temporary events such as concerts and football matches, etc., to reduce the traffic congestion at the existing nearby terrestrial base stations (BSs). Moreover, UAVs can be deployed in order to provide Internet services to the users in the disaster areas where the existing terrestrial networks already collapsed, and to perform search and rescue operations \cite{zeng2019accessing}\cite{zeng2016wireless}. Though UAVs deployment is a solution to extend network coverage of the existing terrestrial networks, there are several challenging issues, e.g., optimal UAVs trajectory and UAVs-users association in order to get good air-to-ground channel quality, efficient communication resources (e.g., subchannels, and power) allocation, and so on \cite{gupta2015survey} \cite{khuwaja2018survey}. The work in \cite{tun2020energy} proposed an efficient UAV trajectory optimization, communication and computation resources allocation framework in the UAV-assisted multi-access edge computing with the aim of minimizing the energy consumption of both UAV and mobile devices.

\subsection{Challenges and Research Contributions}
When we consider resource sharing amongst SPs in the UAVs-assisted wireless networks, it is challenging to ensure efficient resource (i.e., channels) utilization, inter-SP isolation (i.e., no interference among SPs), and intra-SP isolation (i.e., no interference among users in the same SP). Furthermore, unlike terrestrial BSs, UAVs have limited power. Thus, efficient power management is also a significant challenge. To address the above-mentioned challenging issues, in this work, we develop an efficient user association and resource sharing framework in the cell-free UAVs-assisted wireless networks by applying a matching algorithm and a distributed iterative algorithm based on joint whale optimization and Lagrangian relaxation approach. The summary of main contribution of this paper is as follows:

\begin{itemize}
	\item Firstly, we formulate a joint user association and wireless resource sharing problem in the cell-free UAVs networks. Here, we maximize the total utility/profit of SPs while satisfying the QoS constraints of mobile users, and the resource constraints of the UAVs deployed by the MNOs. Then, we notice that the formulated problem is a mixed integer, non-convex problem, which is NP-hard.

	\item We next decompose the formulated problem into two subproblems: users association and resource sharing problems. Then, we deploy two-sided matching algorithm in order to solve the user association problem. Furthermore, whale optimization and Lagrangian relaxation methods are used to address the resource sharing problem. 
	
	\item Finally, we perform extensive simulation in order to demonstrate our proposed solution approach outperforms other benchmark schemes, such as Random channels with optimal power (RCOP), Equal channels with optimal power (ECOP), Random power with optimal channels (RPOC), and Equal power with optimal channels (EPOC), and existing algorithms, namely, Generalized Kelly Mechanism (GKM), Kelly Mechanism (KM), and Equal Sharing (ES). Simulation results show that the total network utility under our proposed algorithm is 2.003\%, 2.06\%, and 8.15\% higher than that of the GKM, KM, and ES, respectively.       
	
	%\item In the simulation results section, we first present the convergence of the proposed algorithm for the joint users association and resource sharing in cell-free UAVs-assisted wireless networks. Then, we compare the achieved utility of each SP under proposed algorithm with 4 benchmark schemes: Random channels with optimal power (RCOP), Equal channels with optimal power (ECOP), Random power with optimal channels (RPOC), and Equal power with optimal channels (EPOC). The proposed algorithm resulted in a substantial performance improvement in comparison to the benchmark schemes. Moreover, we also compare the performance of the proposed algorithm with the existing schemes such as Generalized Kelly Mechanism (GKM), Kelly Mechanism (KM), and Equal Sharing (ES). Simulation results show that the total network utility under our proposed algorithm is 2.003\%, 2.06\%, and 8.15\% higher than that of the GKM, KM, and ES, respectively.                     
        
\end{itemize}

\subsection{Organization}
The rest of this paper is structured as follows: related works are summarized in Section \ref{related}. Section \ref{SMPF} represents the proposed system model and problem formulation. The proposed solution approaches in order to address the formulated problem are presented in Section \ref{solution}. Section \ref{results} demonstrates the numerical results. Finally, we conclude the paper in Section \ref{con}.

\section{Related Works}
\label{related} 
\subsection{Virtualized Wireless Networks}

In \cite{fu2012stochastic}, authors have proposed resource sharing framework in virtualized wireless networks where they modeled the interaction between SPs MNO as a stochastic game. In each resource allocation round, SPs bid for the wireless resources and MNO decides the winning bids. Then, the Vickrey-Clarke-Groves (VCG) mechanism is deployed in order to decide the market clearing price of the resource for the winning SPs. A three-layer game based resource sharing framework for the end users (UEs), mobile virtual network operators (MVNOs), and wireless infrastructure providers (WIPs) in the virtualized wireless networks has been discussed in \cite{rawat2018payoff}. Here, WIPs divide its own radio frequency into multiple slices and lease to MVNOs. Then, MVNOs allow their mobile users to use their subleased frequency slices. In the proposed game, both WIPs and MVNOs choose the best strategies in order to maximize their profits. Furthermore, the work \cite{kazmi2017hierarchical} studied the hierarchical matching game based resource sharing framework in the wireless network virtualization. In \cite{ho2018wireless}, authors has proposed the user clustering, and resource, i.e., resource blocks and power allocation problem in virtualized wireless networks by adopting the non-orthogonal multiple access (NOMA) scheme. Next, the two-sided matching algorithm is applied to cluster the users, and then, the Lagrangian relaxation method is deployed to solve the resource allocation problem. The authors in \cite{kamel2015lte} studied the dynamic radio resource blocks allocation problem in the virtualized multi-cell networks. Then, the fast centralized and heuristic algorithm was proposed to solve the formulated problem. In \cite{gomez2019market}, authors proposed the market driven resource allocation problem in the virtualized wireless networks with the aim of maximizing the revenue of the MNOs while satisfying the QoS constraints of the mobile users. Here, authors considered the mobility of the users and it follows the Poisson point process (PPP). Finally, the matching algorithm is applied in order to solve the proposed problem. Furthermore, a distributed three-sided matching based radio resource allocation framework in the virtualized wireless networks is proposed in \cite{raveendran2019cyclic}. In \cite{han2020hierarchical}, authors studied the hierarchical resource scheduling mechanism in the multi-service networks with the help of wireless network virtualization technology. Here, they divided resource scheduling problem into two-dimension-time scale: large time scale, and small time scale, where inter-slice resource scheduling happens in large time scale and intra-slice resource allocation is in small time scale.

\subsection{UAV-Assisted Wireless Networks}
The work in \cite{hua2019energy} investigated secure data transmission in multi-UAV assisted wireless networks. Moreover, an efficient iterative algorithm has been deployed to solve the formulated problem. A novel framework for joint UAVs' trajectory and power control with the aim of maximizing the sum rate of the mobile users while satisfying the QoS requirement of mobile users has been proposed in \cite{liu2019trajectory}. Then, a three-step solution approach based on the multi-agent Q-learning has been proposed in order to get the mobility of the mobile users and UAVs' trajectory. In \cite{zhang2020predictive}, authors proposed the machine learning based efficient UAVs deployment in order to offload the traffic load of the terrestrial BSs. After that, a contract theory is deployed in order to reinforce the truthful information exchange between the UAVs and ground BSs. The work \cite{liu2019optimization} investigated the joint UAVs deployment, UAVs-users association, wireless backhaul resource allocation problem in the multi-UAV-assisted wireless networks. In \cite{gong2018flight}, authors considered UAV flight time minimization problem in the UAV-assisted sensor network in which UAV collects sensing data from the set of sensor nodes. Moreover, the work of \cite{hua2018power} considered joint sensor scheduling scheme, power allocation, UAV trajectory optimization problem in the sensor network with the aim of minimizing the total energy consumption of the UAV. Then, they applied the block coordinate descent method and successive convex approximation approach in order to solve the problem. Analysis of the outage probability for blockage environment at the millimeter wave (mmWave) band in UAV-aided wireless networks has been discussed in \cite{fontanesi2020outage}. The authors in \cite{rahmati2019dynamic} proposed the dynamic interference management and sum rate maximization problem in the integrated air-ground network where UAVs are deployed as a relay between ground BSs and mobile users. In \cite{oubbati2020softwarization}, authors have considered softwarization in the UAV-aided wireless networks. The work in \cite{xilouris2018uav} proposed network slicing architecture and lightweight virtualization in the UAV networks. In \cite{singhal2019efficient}, the authors introduced software-define networking (SDN) framework to ensure reliable and efficient communication in the UAV-assisted wireless networks with intermittent connectivity and changing network topology according to UAV movement.

All existing studies considered resource sharing amongst SPs in virtualized wireless networks and UAVs-assisted wireless networks independently. Different from existing studies, in this paper, we jointly consider wireless network virtualization technology in UAVs-assisted wireless networks. Then, we formulate the SPs' utility maximization problem in the considered network model by jointly optimizing the users association and resource sharing.

% You must have at least 2 lines in t he paragraph with the drop letter

\begin{figure}[t!]
	\centering
	\captionsetup{justification = centering}
	\includegraphics[width=\linewidth]{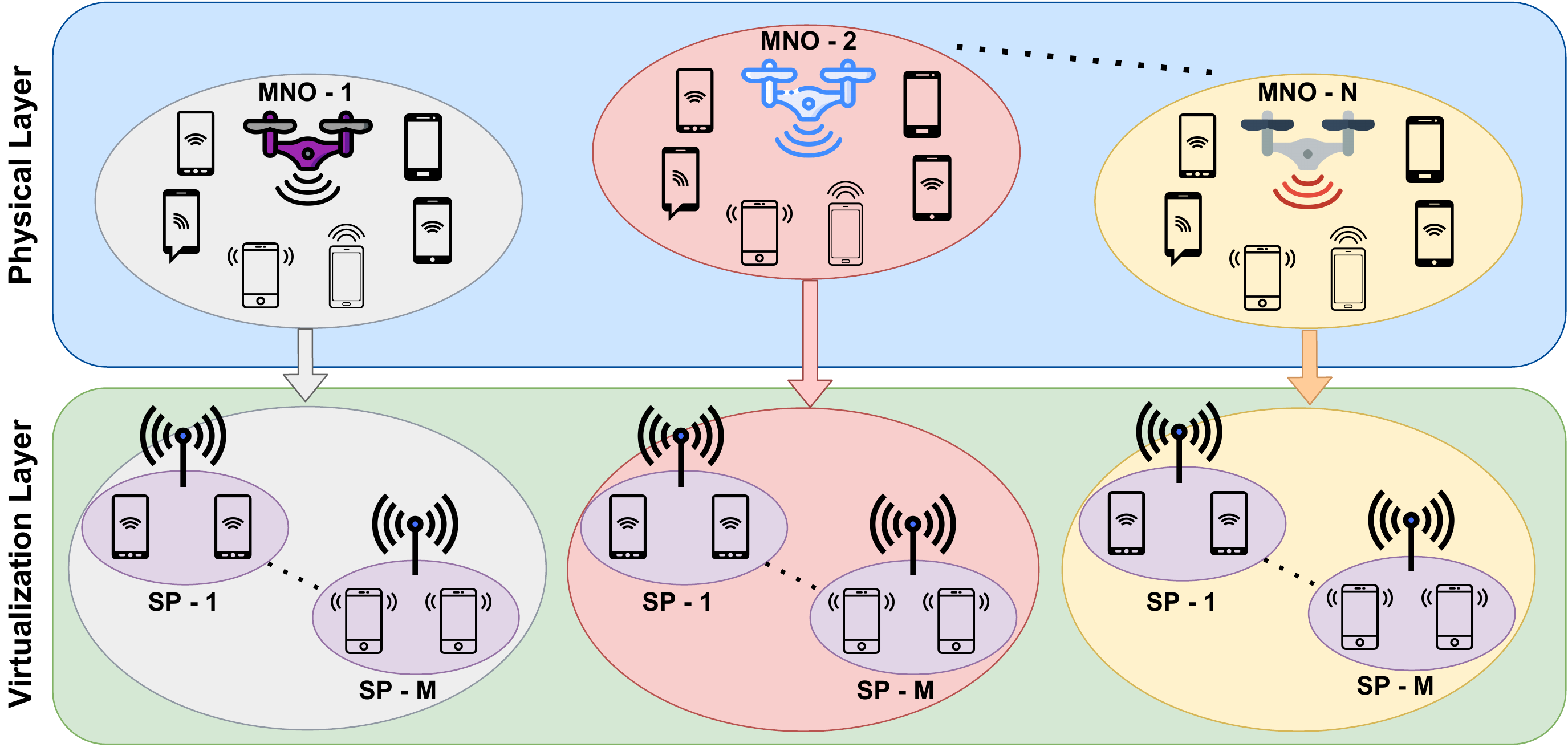}
	\caption{Illustration of our system model.}
	\label{Optimization Framework}
\end{figure}

\begin{table}[t]
	\centering
	\caption{Summary of Notations.}
	\label{tab:table1}
	\begin{tabular}{ll}
		\hline
		Notation & Definition\\
		\hline
		$\mathcal{M}$ & Set of SPs,  $|\mathcal{M}|= M$\\
		$\mathcal{N}$ & Set of MNOs,  $|\mathcal{N}|= N$\\
		$W$           & Total bandwidth at each UAV \\
		$\mathcal{B}_n$ & Set of available subchannels,  $|\mathcal{B}_m|= B_m$  at the UAV \\
	            	 & UAV deployed by MNO $n \in \mathcal{N}$ \\
	    $\mathcal{B}$ & Set of available subchannels, $|\mathcal{B}|= B$ in the considered \\
	         & network \\  
		$\mathcal{U}_m$  & Set of mobile users of SP $m \in \mathcal{M}$, $|\mathcal{U}_m|= U_m$ \\
		
		$\mathcal{U}$    & Total number of mobile users in the considered network \\
		
		$o_u^m$      &  Location of mobile user $u \in \mathcal{U}$ of SP $m \in \mathcal{M}$\\
		
		$c_n$    & Location of the UAV deployed by MNO $n \in \mathcal{N}$\\
		$\textbf{A}$  & UAV-user association matrix \\
		$a_{mu}^{n}$  & UAV-user association variable \\
		$\textbf{X}$  & Subchannels assignment matrix \\
		$x^{nb}_{mu}$ & Subchannels assignment variable\\
		$p^{nb}_{mu}$ & transmit power the UAV deployed by MNO $n \in \mathcal{N}$ on  \\
		            &  subchannel $b \in \mathcal{B}$ assigned to user $u$ of SP$m \in \mathcal{M}$\\		  
		$d_{mu}^n$  & Distance between mobile user $u\in \mathcal{U}$ of SP $m \in \mathcal{M}$  \\
		    & and the UAV deployed by MNO $n \in \mathcal{N}$ \\   
		$g_0$      & Channel gain at the reference distance $d_0 =1 $ m \\           
		$g_{mu}^{nb}$  & Channel gain of the mobile user $u \in \mathcal{U}$ of  \\
		              & SP $m \in \mathcal{M}$\\ 
		              
		$\gamma_{mu}^{nb}$ & Received SINR of mobile user $u \in \mathcal{U}$ of SP $m \in \mathcal{M}$ on \\
		  &  subchannel $b \in \mathcal{B}$ of the UAV deployed by MNO  $n \in \mathcal{N}$\\
		$\alpha$   & Path loss exponent \\
		$\omega$   & Carrier frequency of each subchannel \\
		$R^{nb}_{mu}$ & Achievable data rate of the mobile user $u \in \mathcal{U}$ of SP  \\
		          & $m \in \mathcal{M}$ on subchannel $b\in \mathcal{B}$ of the UAV deployed\\
		          & by MNO $n \in \mathcal{N}$ \\
		$R^{\mathbf{min}}_{mu}$  & Minimum data rate requirement of mobile user $u \in \mathcal{U}$ \\
		  &   of SP $m \in \mathcal{M}$ \\
		$U_m$  & Network utility of SP $m \in \mathcal{M}$ \\
		
		$U_m^{\textrm{Rev}} $ & Revenue of the SP $m \in \mathcal{M}$ \\
		$U_m^{\textrm{Cost}} $ & Total cost of the SP $ m \in \mathcal{M}$ \\
		$\delta_m^u$   & The payment (unit price per Mbps) of mobile user $u \in \mathcal{U}$ \\
		  & to SP $m \in \mathcal{M}$ \\
		  
		$\beta_n^b$  & Unit price per subchannel set by the MNO $n \in \mathcal{N}$ \\
		
		$\theta_n$  & Unit price per a unit of transmit power set by the MNO \\
		& $n \in \mathcal{N}$ \\

		\hline
	\end{tabular}
\end{table}

\section{System Model and Problem Formulation}
\label{SMPF} 
\subsection{Overview System Model}
As illustrated in Fig.~\ref{Optimization Framework}, we consider a UAV-assisted wireless network where a set $\mathcal{M}$ of $M$ SPs rent wireless resources (i.e., sub-channel and power) from a set of MNOs, $\mathcal{N} = \{1,2, \dots, N\}$ with the help of wireless network virtualization technology. For simplicity, we consider each MNO deploys a single UAV embedded with the communication chips to provide connectivity out of the coverage area of the terrestrial network. The UAVs installed by the MNOs are operating on the different frequency bands, and therefore, there will not be inter-cells interference among UAVs. Moreover, following the orthogonal frequency division multiple access (OFDMA) scheme, the total available bandwidth $W$ at each UAV deployed by MVNO $n \in \mathcal{N}$ is further divided into a set of subchannels, $\mathcal{B}_n= \{1,2, \dots, B_n\}$ with the carrier frequency of $\omega$ per subchannel. Therefore,  $\mathcal{B} \triangleq \cup_{n=1}^{|N|} \mathcal{B}_n$ is the total number of subchannels available in the considered network system. Each SP $m \in \mathcal{M}$ is providing specific services such as messaging, video streaming, online gaming, augmented reality (AR) and virtual reality (VR) services, and so on, to their mobile users, $\mathcal{U}_m = \{1,2, \dots, U_m\}$ with the help of radio resources from MNOs. As a result, $\mathcal{U} \triangleq \cup_{m=1}^{|M|} \mathcal{U}_m$ is the total number of users in the system.

\subsection{ UAV-Assisted Wireless Communication Model}
The UAVs deployed by MVNOs are hovering at the fixed location over the users of the SPs in order to provide wireless communication support. Let $o_u^m = [x_u^m, y_u^m]^T$ and $c_n = [x_n, y_n, h_n]^T $ be the horizontal coordinate of the mobile user $u \in \mathcal{U}$ of SP $m\in \mathcal{M}$ and the location of the UAV installed by the MNO $n \in \mathcal{N}$, respectively.

\textbf{Users Association}. Let $ \textbf{A} \in \mathbb{R}^{|\mathcal{U}| \times |\mathcal{N}|}$ be association matrix for all users $|\mathcal{U}|$ in the system over all UAVs deployed by $|\mathcal{N}|$ MNOs. Then, we can describe each element of association matrix as follows:       

\begin{equation}
a_{mu}^{n} =
\begin{cases}
1, \ \ \text{ if user $u$ of SP $m$ is associated to the UAV}\\
\ \ \ \ \ \text{deployed by MNO $n$}, \\
0, \ \ \text{otherwise}.
\end{cases}
\end{equation}
In this work, we assume that the user of each SP can be associated with at most one UAV, i.e., 

\begin{equation}
\sum_{n=1}^{N} a_{mu}^{n} \leq 1,    \forall u \in \mathcal{U}, \forall m \in \mathcal{M}.  
\end{equation} 

\textbf{Subchannels Assignment}. Let $\textbf{X}\in \mathbb{R}^{|\mathcal{U}| \times (|\mathcal{N}| \times |\mathcal{B}|)}$ be the subchannel assignment matrix for all  users $\mathcal{U}$ over all subchannels $|\mathcal{B}|$ of all UAVs deployed by the $|\mathcal{N}|$ MNOs. Then, the element of the subchannels assignment matrix can be introduced as follows:

\begin{equation}
x_{mu}^{nb} =
\begin{cases}
1, \ \ \text{ if user $u$ of SP $m$ is assigned to subchannel $b$}\\
\ \ \  \text{of UAV deployed by MNO $n$}, \\
0, \ \ \text{otherwise}.
\end{cases}
\end{equation}
We also consider that each subchannel cannot can be assigned to more than one mobile user, i.e., 

\begin{equation}
\sum_{m=1}^{M} \sum_{u=1}^{U_m} x_{mu}^{nb} \leq 1,    \forall b \in \mathcal{B}, \forall n \in \mathcal{N}.  
\end{equation} 

\textbf{Downlink Tranmission Rate}. The downlink received signal to noise ratio (SNR) of user $u$ of SP $m$ on subchannel $b$ of the UAV deployed by MNO $n$ as follows: 

\begin{equation} \label{SINR}
\gamma_{mu}^{nb} = \frac{p_{mu}^{nb}g_{mu}^{nb}}{\sigma^2}, \forall u \in \mathcal{U}, \forall m \in \mathcal{M}, \forall b \in \mathcal{B}, \forall n \in \mathcal{N}, 
\end{equation}  
where $p_{mu}^{nb}$ and $g_{mu}^{nb}$ are the channel gain and the transmit power of the UAV deployed by MNO $n$ on subchannel $b$ assigned to user $u$ of SP $m$, respectively, and $\sigma^2$ is additive Gaussian noise power. Denote $\textbf{P}\in \mathbb{R}^{|\mathcal{U}| \times (|\mathcal{N}| \times |\mathcal{B}|)}$ is the transmit power matrix of the UAVs deployed by $|\mathcal{N}|$ MNOs. In this work, we consider that the downlink data transmission from UAV to user is dominated by the line-of-sight (LoS) link, moreover, the free-space path loss model is adopted. Therefore, the achievable channel gain between UAV deployed by MNO $n$ and user $u$ of SP $m$ assigned to subchannel $b$ as follows:

\begin{equation} \label{SINR}
g_{mu}^{nb} = \frac{g_0}{(d_{mu}^n)^\alpha}, \forall u \in \mathcal{U}, \forall m \in \mathcal{M}, \forall b \in \mathcal{B}, \forall n \in \mathcal{N},
\end{equation}  
where $g_0$ is the channel gain at the reference distance $d_0 =1$ m, $\alpha$ is the pass loss exponent. $d_{mu}^n$ is the distance between user $u$ of SP $m$ and UAV of MNO $n$ and it is as follows:

\begin{equation} \label{SINR}
d_{mu}^n  = \sqrt{(h_n^2 + || o_u^m - c_n||^2 )},    \forall u \in \mathcal{U}, \forall m \in \mathcal{M}, \forall n \in \mathcal{N}.   
\end{equation}  

Finally, we can write the achievable data rate of user $u$ of SP $m$ assigned to subchannel $b$ at the UAV of MNO $n$ as follows:
\begin{equation}
R_{mu}^{nb} = \omega \log_2(1 + \gamma_{mu}^{nb} ), \forall u \in \mathcal{U}, \forall m \in \mathcal{M}, \forall b \in \mathcal{B}, \forall n \in \mathcal{N}.
\end{equation}

\textbf{QoS Requirement.} The achievable downlink transmission of each user has to be higher than its minimum rate requirement. Therefore, by mathematically, we can express the required QoS constraint of user $u$ of SP $m$ assigned on subchannel $b$ of the UAV deployed by MNO $n$ as follows: 

\begin{equation}
\sum_{n=1}^{N}\sum_{b=1}^{B_n} a_{mu}^{n} x_{mu}^{nb} R_{mu}^{nb} \geq R_{mu}^{\mathsf{min}},  \forall u\in \mathcal{U}, \forall m \in \mathcal{M}, 
\end{equation}
where $R_{mu}^{\mathsf{min}}$ is the minimum required data rate of user $u$ of SP $m$.    

\textbf{Network Utility Function}. Let $U_m (\textbf{A}, \textbf{X}, \textbf{P})$ be network utility function of SP $m$ in this UAV-assisted wireless network and we can express as follows:
\begin{equation}
U_m (\textbf{A}, \textbf{X}, \textbf{P}) = U_m^{\textrm{Rev}} (\textbf{A}, \textbf{X}, \textbf{P}) - U_m^{\textrm{Cost}}(\textbf{A}, \textbf{X}, \textbf{P}), \forall m \in \mathcal{M},
\end{equation}
where $ U_m^{\textrm{Rev}} (\textbf{A}, \textbf{X}, \textbf{P})$ is the total revenue of SP $m$ from its mobile users $\mathcal{U}_m$, i.e.,
\begin{equation}
 U_m^{\textrm{Rev}} (\textbf{A}, \textbf{X}, \textbf{P}) = \sum_{n=1}^{N} \sum_{b=1}^{B_n} \sum_{u=1}^{U_m} \delta_{m}^u a_{mu}^{n} x_{mu}^{nb} R_{mu}^{nb},  \forall m \in \mathcal{M},
\end{equation} 
where $\delta_m^u$ is the payment (unit price per Mbps) of the mobile user $u$ to SP $m$. Then, $U_m^{\textrm{Cost}}(\textbf{A}, \textbf{X}, \textbf{P})$ is total cost that SP $m$ needs to pay to the MNOs for leasing wireless resources, subchannels and transmit power, i.e.,

\begin{equation}
\begin{split}
U_m^{\textrm{Cost}}(\textbf{A}, \textbf{X}, \textbf{P}) &=  \sum_{n=1}^{N}  \sum_{b=1}^{B_n} \beta_n^b\sum_{u=1}^{U_m} a_{mu}^{n}x_{mu}^{nb} + \sum_{n=1}^{N} \theta_n \sum_{b=1}^{B_n} \sum_{u=1}^{U_m} \\
& \ \ \ \ \ \ \ \ \ \ \ \ \ \ \ \ \ \  \ \ \ \ a_{mu}^{n} p_{mu}^{nb}, \forall m \in \mathcal{M},   
\end{split}  
\end{equation}
where $\beta_n^b$ and $\theta_n$ are the unit price per subchannel and transmit power set by MNO $n$.

\subsection{Problem Formulation}
In this subsection, we will present the detailed problem formulation of the proposed system model. The main objective of this work is to maximize sum of the utility of SPs under the given price of the resources (i.e., subchannels and power) while satisfying the QoS requirement of their mobile users. Therefore, we can pose our optimization problem formally as follows:

	\begin{maxi!}[2]                 % maxi! = maximize
		{\boldsymbol{A, X, P}}                               % optimization variable
		{\sum_{m=1}^{M}  U_m (\textbf{A}, \textbf{X}, \textbf{P})}{\label{opt:P1}}{\textbf{P:}} 
		\addConstraint{\sum_{n=1}^{N}\sum_{b=1}^{B_n} a_{mu}^{n} x_{mu}^{nb}R_{mu}^{nb} \geq R_{mu}^{\mathsf{min}}, \forall u, \forall m \in \mathcal{M},}
		\addConstraint{ \sum_{n=1}^{N} a_{mu}^{n} \leq 1,   \ \ \ \ \ \  \forall u \in \mathcal{U}, \forall m \in \mathcal{M},} 
		\addConstraint{\sum_{m=1}^{M} \sum_{u=1}^{U_m} a_{mu}^{n} \leq Q^n, \ \ \ \  \forall n \in \mathcal{N},}
		\addConstraint{\sum_{m=1}^{M} \sum_{u=1}^{U_m} x_{mu}^{nb} \leq 1,  \ \ \ \forall b \in \mathcal{B}, \forall n \in \mathcal{N},}
		\addConstraint{\sum_{b=1}^{B_n} \sum_{m=1}^{M} \sum_{u=1}^{U_m}    x_{mu}^{nb} p_{mu}^{nb} \leq P_n^{\mathsf{max}},  \ \forall n \in \mathcal{N}, }
    	\addConstraint{0 \leq  p_{mu}^{nb} \leq  P_n^{\mathsf{max}},  \forall u \in \mathcal{U}, \forall m \in \mathcal{M}, \forall b \in \mathcal{B},}
		\addConstraint{a_{mu}^{n} \in \{0,1\}, \ \ \forall u \in \mathcal{U}, \forall m \in \mathcal{M}, \forall n \in \mathcal{N},}
		\addConstraint{x_{mu}^{nb} \in \{0,1\},  \forall u \in \mathcal{U}, \forall m \in \mathcal{M}, \forall n \in \mathcal{N},}
	\end{maxi!}
where constraint (13b) ensures the QoS constraint of each mobile user of each SP, (13c) presents that each mobile user can associate with at most one UAV, (13d) ensures that each UAV serves at most $Q^n$ users in the network, (13e) demonstrates a sub-channel of each UAV cannot be assigned to more than one mobile user, (13f) and (13g) assure the total transmit power of the UAV on the channels to the users of SPs have to be less than the maximum transmit power of the UAV. Finally, (13h) and (13i) are the binary constraints of the UAV-user association and subchannel assignment.

\section{Proposed Solution Approaches for Formulated Problem}
\label{solution} 

We can see from (13) that our formulated problem is a mixed-integer non-linear (MINL) problem. In other words, it is an NP-hard problem, and therefore, it is not possible to get solution within polynomial time. Hence, we decompose the problem into two subproblems: 1) users association problem and 2) resource sharing problem in the following subsections.

\subsection{Two-Sided Matching-Based Users Association}
%\label{solution} 

We assume that the total transmit power of the UAV deployed by MNO $n \in \mathcal{N}$ is divided equally among the set of subchannels $\mathcal{B}_n$. Then, the user $u \in \mathcal{U}$ is presumed to be oblivious to all the subchannels provided by that UAV. This allows the power and subchannel assignment variables to be ignored from our proposed problem in (13). Therefore, we can reformulate the problem in (13) as a users association problem and it is as follows:
	\begin{maxi!}[2]                 % maxi! = maximize
		{\boldsymbol{A}}                               % optimization variable
		{\sum_{m=1}^{M}  U_m (\textbf{A})}{\label{opt:P1}}{\textbf{P1:}} 
		\addConstraint{\sum_{n=1}^{N}\sum_{b=1}^{B_n} a_{mu}^{n} R_{mu}^{nb} \geq R_{mu}^{\mathsf{min}}, \forall u, \forall m \in \mathcal{M},}
		\addConstraint{ \sum_{n=1}^{N} a_{mu}^{n} \leq 1,   \ \ \ \ \ \  \forall u \in \mathcal{U}, \forall m \in \mathcal{M},} 
		\addConstraint{\sum_{m=1}^{M} \sum_{u=1}^{U_m} a_{mu}^{n} \leq Q^n, \ \ \ \  \forall n \in \mathcal{N},}
		\addConstraint{a_{mu}^{n} \in \{0,1\}, \ \ \forall u \in \mathcal{U}, \forall m \in \mathcal{M}, \forall n \in \mathcal{N}}.
	\end{maxi!}
Then, in order to get the tractable solution, the users association problem in (14) can be modeled as a two-sided matching game \cite{roth1992two}. In this game, there are two disjoint sets of players: the set of users $\mathcal{U}$, and the set of UAV $\mathcal{N}$. In our proposed matching game, each user $u \in \mathcal{U}$ can be associated with at most one UAV. Though, each UAV can serve a certain number of users, which depends on the number of available subchannels and the maximum number of allowable users $Q^n$ at UAV $n \in \mathcal{N}$. Therefore, our model refers to a \emph{one-to-many matching} given by the tuple $(\mathcal{U}, \mathcal{N},  Q^n, \succ_{\mathcal{U}}, \succ_{\mathcal{N}})$ where  $\succ_{\mathcal{U}} \triangleq\{\succ_{u_m}\}_{u_m\in\mathcal{U}}$ and $\succ_{\mathcal{N}} \triangleq \{ \succ_n\}_{n \in \mathcal{N}}$ states the set of preference relations between users of SPs and UAVs.  
\textbf{Definition 1.} \textit{A matching $\vartheta$ is defined as the function from the set $\mathcal{U} \ \cup \ \mathcal{N}$ into the set of $\mathcal{U} \ \cup \ \mathcal{N}$ such that:\\ \\ 
(1)  $|\vartheta(u_m)| \leq 1$ and $\vartheta(u_m) \in \mathcal{N}$, \\
(2)  $|\vartheta(n)| \leq Q^n$ and $\vartheta (n) \in 2^{|Q^n|} \cup \emptyset$, \\
(3)  $\vartheta(u_m) = n$ if and only if $u_m$ is in $\vartheta(n)$}, \\ \\
where $|\vartheta(\cdot)|$ represents the cardinality of the matching outcome $\vartheta(\cdot)$. Moreover, the first two conditions of the above-mentioned definition satisfy constraints (14c) and (14d).

\textbf{Preference Lists of Players}. Each user $u_m \in \mathcal{U}$ of SP $m\in \mathcal{M}$ calculates the achievable utility and data rate with each UAV $n\in \mathcal{N}$, then, sorts UAVs which have data rate greater than the required data rate in a decreasing order in order to construct his/her preference list. Let $\mathcal{P}_{u_m}$ be the preference list of SP's user $u_m$, and it can be demonstrated by the vector of utility function. Mathematically, it is as follows:
\begin{equation}
    T_{u_m}(n) = [U_{mu_m}^n = U_{mu_m}^{\textrm{Rev}} - U_{mu_m}^{\textrm{Cost}} ]_{n \in \mathcal{N}, R_{mu}^{nb} \geq R_{mu}^{\mathsf{min}}}.     
\end{equation}
On this spot, the user of each SP desires in order to associate with the UAV, and so it can achieve the maximum utility function. For instance, $n\succ_{u_m}n'$ represents that mobile user $u_m$ of SP $m \in \mathcal{M}$ prefers to associate with UAV $n$ as opposed to UAV $n'$, i.e., $T_{u_m}(n) > T_{u_m}(n')$. Similarly, to construct the preference profile $\mathcal{P}_n$, each UAV $n$ determines the achievable data rate with each SP's user $u_m$ and ranks in a decreasing order. The preference of UAV $n$ can be expressed as the vector of its utility function and it is as follows:        
\begin{equation}
    T_{n}(u_m) = \big[R_{mu_m}^{n} = \omega \log_2(1+\gamma_{mu_m}^{nb})\big]_{u_m\in\mathcal{U}}, 
\end{equation}

Our goal for the formulated two-sided matching game described is to find a stable matching, which is a key solution concept. To get the stable matching, there should be no blocking pair. Therefore, we formally define the stable matching of the formulated two-sided matching game as follows:      

%\textbf{Definition 2.} \textit{A matching $\vartheta$ is stable if there is no blocking pair $(u_m, n)$, where $u_m \in \mathcal{U}, n \in \mathcal{N}$, such that $\mathcal{Q}^n_{\textrm{sur}} \geq \mathcal{Q}^n_{u_m}$, $u_m \succ_n \emptyset$, and $n \succ_{u_m} \vartheta(u_m)$, where $\vartheta(u_m)$ demonstrates the existing matched partners of $u_m$.} \\

\textbf{Definition 2.} \textit{A matching $\vartheta$ is stable if there is no blocking pair $(u_m, n)$, where a pair $(u_m, n)$ is blocking when $u_m \notin \vartheta(n), n \notin \vartheta(u_m)$, and $n\succ_{u_m}\vartheta(n)$ and $u_m\succ_{n}\vartheta(u_m)$}. \\

%Here, $\mathcal{Q}^n_{\textrm{sur}} = \mathcal{Q}^n - \sum{u_m \in \vartheta(n)} \mathcal{Q}^n_{u_m}$ represents the remaining quota at the UAV $n$. The quota of the UAV $n \in \mathcal{N}$ is filled when $ \mathcal{Q}^n_{\textrm{sur}} < \mathcal{Q}^n_{u_m} $ for a requesting $u_m$ \in \mathcal{U}.
Algorithm \ref{alg:1} describes the matching-based users association problem. The users association problem is formulated as a many-to-one matching game, as previously stated. To that end, our objective is to derive a stable matching, which is the outcome of deferred-acceptance algorithm \cite{gale1962college}. The process starts with building preference profiles, i.e., $\mathcal{P}_{u_m}$ for user $u_m$ and $\mathcal{P}_n$ for UAV $n$, while in each iteration,
each user $u_m$ opt to highest preferred UAV for association, as presented in line 5 of Algorithm \ref{alg:1}. On receiving the proposal from user $u_m$, each UAV $n$  temporarily accepts $Q^n$ proposals of users having the highest ranked in UAV $n$'s preference list and  rejects the other lower ranked proposals as shown by lines 6–8 of Algorithm \ref{alg:1}. However, if UAV $n$ prefers the new proposal to the current one, as demonstrated by lines 9-12 of Algorithm \ref{alg:1}, it will reject the existing proposal and accept the new one. The users with least preference profile  ${u'}_{m'} \in S_n[t]$ get rejected and dropped from the preference list of each UAV $P_n[t]$. Correspondingly, these users also extract UAV $n$ from their preference list $P_{{u'}_{m'}}[t]$ (line 13-15). The iterative approach of rejection, and deference (line 13-15) continues unless all users have received a satisfactory proposal, at which stage a stable solution to the users association problem is obtained. Finally, the output of the many-to-one matching $\vartheta$, is transformed to drive the users association vector $\boldsymbol{A}$ for the problem (line 17), i.e., $\vartheta \rightarrow \boldsymbol{A}$. The complexity of two-sided matching game based users association algorithm depends on the required number of accepting/rejecting decisions to attain the stable matching $\vartheta$. In each iteration of Algorithm 1, each user in the network proposes to associate with the most preferred UAV in their preference list, and then the UAV determines whether to accept or reject the proposal. Here, the maximum size of the preference list of each user is $|\mathcal{N}|$. As a result, Algorithm 1 converges to the stable matching in $\mathcal{O}(|\mathcal{U} \times \mathcal{N}|)$ iterations, in any case of a matching problem \cite{gao2019licensed}, where $\mathcal{U}$ and $ \mathcal{N}$ are the number of users and UAVs in the considered network.

\begin{algorithm}[t!]
	\caption{\strut Two-Sided Matching-Game Based Users Association} 
	\label{alg:1}
	\begin{algorithmic}[1]
	   \STATE{\textbf{Input:} $\mathcal{P}_{u_m}$, $\mathcal{P}_n$, $\forall u_m \in \mathcal{U}, \forall n \in \mathcal{N}$} 
		\STATE{\textbf{Initialize:} $t=0$; $\vartheta[0] \triangleq \{\vartheta(n)[0], \vartheta(u_m)[0]\}_{\forall n, u_m} = \emptyset $ \\ 
		 $\mathcal{S}_n[0] = \emptyset$,  $\mathcal{P}_{u_m}[0] = \mathcal{P}_{u_m}, \forall u_m \in \mathcal{U}$, $\mathcal{P}_{n}[0]= \mathcal{P}_{n}, \forall n \in \mathcal{N}$}, 
		
		\REPEAT
		\STATE{$t \leftarrow t +1$};
		\STATE{\textbf{for} $u_m\in \mathcal{U}$, choose $n \in \mathcal{N}$ depending on the preference list $\mathcal{P}_{u_m}[t]$ \textbf{do}};
		\STATE{\ \ \  \textbf{while} $ u_m \notin \vartheta(n)[t]$  \textbf{do}}
		\STATE{ \ \ \ \ \ \ \ \ \  \textbf{if} $ |\vartheta({n})[t]| < Q^n$ \textbf{then}}
		\STATE{\ \ \ \ \ \ \ \ \ \ \ \ \ $\vartheta(n)[t] = \vartheta(n)[t] \cup u_m$};
		\STATE{\ \ \ \ \ \ \ \ \  \textbf{else if} $|\vartheta({n})[t]| = Q^n$ \text{and} $u_m \succ_{n} \vartheta({n})[t]$ \textbf{then}}
	    \STATE{ \ \ \ \ \ \ \ \ \ \ \ \ \ $\vartheta(n)[t] \leftarrow \vartheta(n)[t] \setminus {u'}_{m'}$};
		\STATE{\ \ \ \ \ \ \ \ \ \ \ \ \ $\vartheta(n)[t] = \vartheta(n)[t] \cup u_m$};
		\STATE{\ \ \ \ \ \ \ \ \ \ \ \ \ $\mathcal{S}_{n}[t] =\{{u'}_{m'} \in \vartheta(n)[t]| u_m\succ_{n}{u'}_{m'}\}$  };
		\STATE{\ \ \ \ \ \ \ \ \  \textbf{for} $s \in \mathcal{S}_{n}[t] \textbf{do}$};
		\STATE{\ \ \ \ \ \ \ \ \ \ \ \ $\mathcal{P}_s[t] \leftarrow  \mathcal{P}_s[t] \setminus {n}$ };
		\STATE{\ \ \ \ \ \ \ \ \ \ \ \ $\mathcal{P}_{n}[t] \leftarrow  \mathcal{P}_{n}[t] \setminus {s}$ };
		\UNTIL{ $ \vartheta[t] = \vartheta[t-1]$};
		\STATE{\textbf{Users Association:} $\vartheta \rightarrow \boldsymbol{A} $.}
		%\STATE{\textbf{Users Association:} \vartheta \rightarrow \boldsymbol{A}}.
    \end{algorithmic}
	\label{Algorithm}
\end{algorithm}

\subsection{Iterative Algorithm-Based Resource Sharing}

At the fixed users association, we can rewrite the resource sharing problem amongst SPs' users as follows:

	\begin{maxi!}[2]                 % maxi! = maximize
		{\boldsymbol{ X, P}}                               % optimization variable
		{\sum_{m=1}^{M}  U_m (\textbf{X}, \textbf{P})}{\label{opt:P1}}{\textbf{P2:}} 
		\addConstraint{\sum_{n=1}^{N}\sum_{b=1}^{B_n}x_{mu}^{nb}R_{mu}^{nb} \geq R_{mu}^{\mathsf{min}}, \forall u, \forall m \in \mathcal{M},}
		\addConstraint{\sum_{m=1}^{M} \sum_{u=1}^{U_m} x_{mu}^{nb} \leq 1,  \ \ \ \forall b \in \mathcal{B}, \forall n \in \mathcal{N},}
		\addConstraint{\sum_{b=1}^{B_n} \sum_{m=1}^{M} \sum_{u=1}^{U_m}    x_{mu}^{nb} p_{mu}^{nb} \leq P_n^{\mathsf{max}},  \ \forall n \in \mathcal{N}, }
    	\addConstraint{0 \leq  p_{mu}^{nb} \leq  P_n^{\mathsf{max}},  \forall u \in \mathcal{U}, \forall m \in \mathcal{M}, \forall b \in \mathcal{B},}
		\addConstraint{x_{mu}^{nb} \in \{0,1\},  \forall u \in \mathcal{U}, \forall m \in \mathcal{M}, \forall n \in \mathcal{N},}
	\end{maxi!}
From (15), we observe that the variables $\boldsymbol{X}$, and $\boldsymbol{P}$ are coupling in both objective function and constraints. Moreover, (15f) is a binary variable. Therefore, our proposed resource sharing problem in (15) is MINL problem, i.e., non-convex problem and it is challenging to solve. Even though the problem is a non-convex problem, at the given subchannels assignment scheme, the power allocation problem becomes convex, and vice versa. Therefore, we decompose the resource sharing problem into two subproblems: subchannels assignment problem and power allocation problem. Then, the two subproblems are solved iteratively.

\subsection{Whale Optimization Based Subchannels Assignment at Given Power Allocation }

At a given power allocation, we can formulate subchannls assignment problem as follows:  

	\begin{maxi!}[2]                 % maxi! = maximize
		{\boldsymbol{X}}                               % optimization variable
		{\sum_{m=1}^{M}  U_m (\textbf{X})}{\label{opt:P1}}{\textbf{P21:}} 
		\addConstraint{\sum_{n=1}^{N}\sum_{b=1}^{B_n}x_{mu}^{nb}R_{mu}^{nb} \geq R_{mu}^{\mathsf{min}}, \forall u, \forall m \in \mathcal{M},}
		\addConstraint{\sum_{m=1}^{M} \sum_{u=1}^{U_m} x_{mu}^{nb} \leq 1,  \ \ \ \forall b \in \mathcal{B}, \forall n \in \mathcal{N},}
		\addConstraint{x_{mu}^{nb} \in \{0,1\},  \forall u \in \mathcal{U}, \forall m \in \mathcal{M}, \forall n \in \mathcal{N}}. 
	\end{maxi!}

The decomposed problem in (18) is a non-convex and combinatorial problem. Thus, to solve the decomposed problem (18), we deploy whale optimization algorithm (WOA) \cite{mirjalili2016whale}. WOA is a meta-heuristic algorithm inspired by the prey hunting behavior of whales. WOA features adaptive mechanisms that balance this algorithm's exploration and exploitation characteristics. In comparison to other heuristic methods, this enhances the likelihood of avoiding local optimum solutions. Furthermore, as WOA is simple to use and adaptable, it may be used to a wide range of optimization problems. The whale's behavior is divided into three, and according to the behavior, we can approach the optimal root.

\textbf{Encircling Prey}. When whales perform in this action, they first evaluate their prey's position before totally engulfing them. The current best whale is thought to be quite close to the optimal solution. The position of the other whales is updated based on the best whale's position. The following equations can describe the behavior:

\begin{equation} \label{D_ep}
\Vec{D} = \left\vert \Vec{C}\cdot\overrightarrow{X^*}(t)-\Vec{X}(t) \right\vert,
\end{equation}

\begin{equation} \label{X_next_ep}
\Vec{X}(t+1) = \lfloor \overrightarrow{X^*}(t) - \Vec{A} \cdot \Vec{D} \rfloor,
\end{equation}
where $t$ is the current iteration, $\left\vert \cdot \right\vert$ is the absolute value, $\lfloor \cdot \rfloor$ is the floor function for the discrete space of the channel, and $\cdot$ denotes the element-wise multiplication. $\Vec{C}=2\cdot\Vec{r}$ and $\Vec{A} = 2\Vec{a}\cdot\Vec{r}-\Vec{a}$ are coefficient vectors in which $\Vec{a}$ is linearly decreased from $2$ to $0$ over iterations, and $\Vec{r}$ is a random vector in the range $[0,1]$. \\

\textbf{Bubble-Net Attacking Method}. After that, a spiral path with a helix shape is formed to simulate the movement of humpback whales, which can be expressed as follows:

\begin{equation} \label{D_bn}
\Vec{D^\prime} = \left\vert \overrightarrow{X^*}(t)-\Vec{X}(t) \right\vert,
\end{equation}

\begin{equation} \label{X_next_bn}
\Vec{X}(t+1) = \lfloor \overrightarrow{D^\prime} \cdot e^{bl} \cdot \cos{(2 \pi l)} + \overrightarrow{X^*}(t) \rfloor,
\end{equation}
where $b$ is a constant value used to determine the logarithmic spiral's shape, and $l$ is a random number in the range $[-1,1]$. \\ 

\textbf{Search for Prey}. The exploration is utilized to explore the global optimum by randomly selecting a position vector $X_{\mathrm{rand}}$ from the present position to create the random search for the prey. The following is a description of the model:

\begin{equation} \label{D_sp}
\Vec{D} = \left\vert \Vec{C}\cdot\overrightarrow{X_{\mathrm{rand}}}(t)-\Vec{X}(t) \right\vert,
\end{equation}

\begin{equation} \label{X_next_sp}
\Vec{X}(t+1) = \lfloor \overrightarrow{X_{\mathrm{rand}}}(t) - \Vec{A} \cdot \Vec{D} \rfloor.
\end{equation} \\

\begin{algorithm}[t!]
	\caption{\strut WOA based Subchannels Assignment at Given Power Allocation} 
	\label{alg:profit}
	\begin{algorithmic}[1]
	    \STATE{\textbf{Input:} the current subchannel assignment $\boldsymbol{X_0}$, the given power allocation $\boldsymbol{P}$.} 
		\STATE{\textbf{Initialize:} the whale population $X_{i}$, $i=\left\{ 1,...,K \right\}$, iteration $t=1$, maximum number of iterations $I_{\mathsf{max}}$.}
		\STATE{Calculate the fitness of the search agents $X_i$ by \eqref{Fitness} and set the best search agent $\overrightarrow{X^*}(0)$}.
		\REPEAT
		\STATE{\textbf{for} $k \leftarrow 1$ to $K$ (the number of whales) \textbf{do}}
		\STATE{\ \ \ Update $a,A,C,l$ and $p$.}
		\STATE{\ \ \ \textbf{if} $p < 0.5$ \textbf{then}}
		\STATE{\ \ \ \ \ \ \textbf{if} $\left| A \right| < 1$ \textbf{then}}
		\STATE{\ \ \ \ \ \ \ \ \ Update $\Vec{D}$ by \eqref{D_ep} and $\Vec{X}$ by \eqref{X_next_ep}.}
		\STATE{\ \ \ \ \ \ \textbf{else}}
		\STATE{\ \ \ \ \ \ \ \ \ Select a random $\overrightarrow{X_{\mathrm{rand}}}$ and update $\Vec{D}$ by \eqref{D_sp}.}
		\STATE{\ \ \ \ \ \ \ \ \ Update the position $\Vec{X}$ by \eqref{X_next_sp}.}
		\STATE{\ \ \ \textbf{end if}}
		\STATE{\textbf{else}}
		\STATE{\ \ \ Update $\Vec{D}$ by \eqref{D_bn} and $\Vec{X}$ by \eqref{X_next_bn}.}
		\STATE{\textbf{end if}}
		\STATE{\textbf{end for}}
		\STATE{Calculate the fitness of each search agent by \eqref{Fitness}.}
		\STATE{Update $X^{*}(t)$ of the best search agent.}
		\STATE{$t \leftarrow t +1$}
		\UNTIL{$t > I_{\mathsf{max}}$}
		\STATE{\textbf{Output:} The best Subchannels Assignment $\boldsymbol{X^*}$.}
	\end{algorithmic}
	\label{Algorithm}
\end{algorithm}

\textbf{Fitness function for the constraint}. We must use the efficient constraint-handling algorithms to tackle proposed constrained problems because the original WOA was designed for unconstrained optimization \cite{pham2020whale}. The fitness function that was used to choose the optimal search agent is:

\begin{equation} \label{Fitness}
\mathrm{Fitness}(\boldsymbol{X}) = \sum_{m=1}^{M}  U_m (\textbf{X}) - \xi \sum_{n=1}^{N} F_{n}(f_{n}(\textbf{X}))f_{n}^{2}(\textbf{X}),
\end{equation}
\begin{equation} \label{f_n}
f_{n}(\textbf{X}) = \sum_{b=1}^{B_n}x_{mu}^{nb}R_{mu}^{nb} - R_{mu}^{\mathsf{min}}, \forall n\in\mathcal{N}. 
\end{equation}
In \eqref{Fitness}, $\xi$ and $F_{n}(f_{n}(\boldsymbol{X}))$ are penalty factors and index function that $F_{n}(f_{n}(\boldsymbol{X}))=0$ if $f_{n}(\boldsymbol{X}) \geq 0$ and $F_{n}(f_{n}(\boldsymbol{X}))=1$ if $f_{n}(\boldsymbol{X}) < 0$. Algorithm 2 depicts the pseudocode for the WOA algorithm. Computing fitness has a computational complexity of  $\mathcal{O}(ND)$, where $N$ represents the  population of whale and $D$ is the dimension of search agents. Then, the complexity of updating the position vector of all the search agents at each iteration is $\mathcal{O}(ND)$. As a result, the complexity of Algorithm 2 can be represented as $\mathcal{O}(NDI)$, in which $I$ is the number of maximum iterations/generations.

\subsection{Lagrangian Relaxation-Based Power Allocation at Given Subchannels Assignment Scheme}

For problem (15) with fixed subchannels assignment scheme, the transmit power allocation problem can be expressed as follows: 
	\begin{maxi!}[2]                 % maxi! = maximize
		{\boldsymbol{P}}                               % optimization variable
		{\sum_{m=1}^{M}  U_m (\textbf{P})}{\label{opt:P1}}{\textbf{P22:}} 
		\addConstraint{\sum_{n=1}^{N}\sum_{b=1}^{B_n}x_{mu}^{nb}R_{mu}^{nb} \geq R_{mu}^{\mathsf{min}},}
		\addConstraint{\sum_{b=1}^{B_n} \sum_{m=1}^{M} \sum_{u=1}^{U_m}    x_{mu}^{nb} p_{mu}^{nb} \leq P_n^{\mathsf{max}},  \ \forall n \in \mathcal{N}, }
    	\addConstraint{0 \leq  p_{mu}^{nb} \leq  P_n^{\mathsf{max}},  \forall u \in \mathcal{U}, \forall m \in \mathcal{M}, \forall b \in \mathcal{B}.}
\end{maxi!}

\textbf{Lemma 1.} \textit{At the given subchannels assignment scheme, the optimization problem in (\ref{opt:P1}) is a convex problem.} \\ 	

\begin{IEEEproof}
The first order derivative of the objective function in (16a) with respect to $P_{mu}^{nb}$ is as follows:
\begin{equation}
\begin{split}
 \frac{\partial U_m(\boldsymbol{P})}{\partial p_{mu}^{nb}} =& \left( \frac{\delta_{m}^ux_{mu}^{nb}\omega g_0}{\ln(2)(d_{mu}^n)^{\alpha} \sigma^2\left[ 1+ \frac{p_{mu}^{nb}g_0}{(d_{mu}^n)^{\alpha} \sigma^2}  \right]}\right) - \theta_n \\
 & \ \ \ \forall n \in \mathcal{N}, \forall b \in \mathcal{B}, \forall m \in \mathcal{M}, \forall u \in \mathcal{U}.
 \end{split} 
\end{equation}
Then,
\begin{equation}
     \begin{split}
     \frac{\partial^2 U_m(\boldsymbol{P})}{\partial{(p_{mu}^{nb})}^2} =& - \frac{\delta_{m}^u x_{mu}^{nb}\omega g_0^2}{\ln(2) \left[(d_{mu}^n)^{\alpha} \sigma^2+ p_{mu}^{nb}g_0  \right]^2}, \forall n \in \mathcal{N}, \forall b \in \mathcal{B},\\ 
     & \ \ \ \ \ \ \ \ \ \ \ \ \ \ \ \ \ \ \ \ \ \ \ \ \ \ \  \ \ \ \ \ \ \     \forall m \in \mathcal{M}, \forall u \in \mathcal{U},
     \end{split} 
\end{equation}
From (18), it is clear that $\frac{\partial^2 U_m(\boldsymbol{P})}{\partial{(p_{mu}^{nb})}^2} < 0$. Therefore, (16a) is a concave function. Moreover, (16b) and (16c) are convex and linear constraints, respectively. Finally, constraint (16d) is affine. Thus, we can conclude that (16) is a convex problem. 
\end{IEEEproof}
  
Here, we introduce non-negative Lagrangian multipliers,  $\lambda_m^u$, $\mu_n$, and $\nu_{mu}^{nb}$ for constraints (16b), (16c), and (16d), respectively. Then, by integrating the objective function in (16a), and constraints (16b), (16c), (16d), the Lagrangian function of (16) can be expressed as follows:

\begin{equation}
\begin{split}
&\mathcal{L}(\boldsymbol{P}, \boldsymbol{\lambda}, \boldsymbol{\mu}, \boldsymbol{\nu}) =\sum_{n=1}^{N} \sum_{b=1}^{B_n} \sum_{m=1}^{M} \sum_{u=1}^{U_m} \delta_{m}^u x_{mu}^{nb}\omega \log_2\left(1+ \gamma_{mu}^{nb} \right) \\
&- \left(  \sum_{n=1}^{N}  \sum_{b=1}^{B_n} \beta_n^b \sum_{m=1}^{M}\sum_{u=1}^{U_m} x_{mu}^{nb} + \sum_{n=1}^{N} \theta_n \sum_{b=1}^{B_n} \sum_{m=1}^{M} \sum_{u=1}^{U_m} p_{mu}^{nb} \right)
 \\
& + \sum_{m=1}^{M}\sum_{u=1}^{U_m} \lambda_m^u\left(\sum_{n=1}^{N}\sum_{b=1}^{B_n}x_{mu}^{nb} R_{mu}^{nb}- R_{mu}^{\mathbf{min}}  \right) + \sum_{n=1}^{N} \sum_{b=1}^{B_n} \sum_{m=1}^{M} \\
& \sum_{u=1}^{U_m}\nu_{mu}^{nb} 
\bigg(P_n^{\mathbf{max}} - p_{mu}^{nb}\bigg)
+ \sum_{n=1}^{N} \mu_n \bigg( P_n^{\mathsf{max}} - \sum_{b=1}^{B_n} \sum_{m=1}^{M} \sum_{u=1}^{U_m}   \\ & x_{mu}^{nb} p_{mu}^{nb}\bigg)
\end{split} 
\end{equation}
where $\boldsymbol{\lambda} = \left[\lambda_m^u\right]_{1\times(MU)}$, $\boldsymbol{\mu} = \left[ \mu_n\right]_{1\times N}$ and $\boldsymbol{\nu} = \left[ \nu_{mu}^{nb}\right]_{1 \times (NBMU)}$, respectively. The dual problem of $\textbf{P21}$ is formulated as follows:

\begin{align}
\underset{(\boldsymbol{\lambda} \geq 0, \boldsymbol{\mu} \geq 0, \boldsymbol{\nu} \geq 0)}{\min} \
& D( \boldsymbol{\lambda}, \boldsymbol{\mu}, \boldsymbol{\nu}),  
\end{align}
where
\begin{equation}
\begin{split}
   D(\boldsymbol{\lambda}, \boldsymbol{\mu}, \boldsymbol{\nu}) = \ &  \underset{\boldsymbol{P}}{\max} \ \mathcal{L}(\boldsymbol{P}, \boldsymbol{\lambda}, \boldsymbol{\mu}, \boldsymbol{\nu}) \\
   & \text{subject to} \ \ \text{(16b)}, \text{(16c)}, \ \text{and} \ \text{(16d)}.
\end{split}
\end{equation}
As shown above, problem (16) is a convex problem, and thus, there exists a strictly feasible point so that the Slater's condition holds, resulting in strong duality \cite{boyd2004convex}. Therefore, we can solve the problem in (16) via the dual problem of (20). The dual problem of (20) can be solved by using the sub-gradient method in which the dual variables are updated as follows:

\begin{algorithm}[t!]
	\caption{\strut Lagrangian Relaxation-Based Power Allocation}
	\label{alg:profit}
	\begin{algorithmic}[1]
	   \STATE{\textbf{Input:} $\delta_m^u$, $\theta_n$, $\boldsymbol{X}$} 
		\STATE{\textbf{Initialize:} $t=0$; $P_{mu}^{nb}(0)$, $\epsilon > 0$, $ \lambda_m^u (0), \mu_n(0), \nu_{mu}^{nb}(0)> 0$, \text{and} $\varsigma_i(0) >0, (i=1,2,3)$}, 
		
		\REPEAT
		\STATE{$t \leftarrow t +1$};
		\STATE{Update $\varsigma_i(t+1),(i=1,2,3)$ according to (25)};
		\STATE{Update $\lambda_m^u (t+1)$, $\mu_n(t+1)$, $\nu_{mu}^{nb}(t+1)$ according to (22), (23), and (24)};
		\STATE{Update $P_{mu}^{nb}(t+1)$ according to (26)};
		\UNTIL{ $|P_{mu}^{nb}(t+1)-P_{mu}^{nb}(t)| \leq \epsilon$};
		
		\STATE{Then, set $P_{mu}^{nb}(t+1)$ as the desired solution}.
		
	\end{algorithmic}
	\label{Algorithm}
\end{algorithm}
\begin{equation}
    \lambda_m^u(t+1) = \left[ \lambda_m^u(t) - \varsigma_1(t) \left(\sum_{n=1}^{N}\sum_{b=1}^{B_n}x_{mu}^{nb} R_{mu}^{nb}- R_{mu}^{\mathbf{min}}\right) \right]^{+},
\end{equation}

\begin{equation}
    \mu_n(t+1) = \left[ \mu_n(t)- \varsigma_2(t)  \left( P_n^{\mathsf{max}} - \sum_{b=1}^{B_n} \sum_{m=1}^{M} \sum_{u=1}^{U_m}    x_{mu}^{nb} p_{mu}^{nb}\right)\right]^{+},
\end{equation}

\begin{equation}
    \nu_{mu}^{nb}(t+1) = \bigg[ \nu_{mu}^{nb}(t)- \varsigma_3(t)  \bigg( P_n^{\mathsf{max}} - p_{mu}^{nb}\bigg)\bigg]^{+},
\end{equation}
where $\varsigma_i(t), (i= 1, 2, 3)$ are the step sizes which can be derived as follows:

\begin{equation}
    \varsigma_i(t) = \frac{z}{\sqrt{t}}, z >0, i = 1,2,3. 
\end{equation}

\textbf{Proportion 1}. \textit{Based on the Karush-Kuhn-Tucker (KKT) conditions \cite{boyd2004convex}, the optimal power allocation of problem \textbf{P21} is as follows:}

\begin{equation}
    p_{mu}^{nb*}= \left[ \frac{x_{mu}^{nb}\omega(\delta_m^u  + \lambda_m^u)}{\theta_n + \mu_n x_{mu}^{nb} + \nu_{mu}^{nb}} - \frac{(d_{mu}^n)^{\alpha}\sigma^2}{g_0}\right]^{+}.
\end{equation}

\begin{IEEEproof}
 The first order derivative of the Lagrangian function in (19) with respect to $P_{mu}^{nb}$ is as follows:
 \begin{equation}
\begin{split}
\frac{\partial \mathcal{L}(\boldsymbol{P}, \boldsymbol{\lambda}, \boldsymbol{\mu}, \boldsymbol{\nu})}{\partial p_{mu}^{nb}} = \left( \frac{\delta_{m}^ux_{mu}^{nb}\omega g_0}{\ln(2)(d_{mu}^n)^{\alpha} \sigma^2\left[ 1+ \frac{p_{mu}^{nb}g_0}{(d_{mu}^n)^{\alpha} \sigma^2}  \right]}\right) - \theta_n   \\  + \lambda_m^u\left(\frac{x_{mu}^{nb}\omega g_0}{\ln(2)(d_{mu}^n)^{\alpha} \sigma^2\left[ 1+ \frac{p_{mu}^{nb}g_0}{(d_{mu}^n)^{\alpha} \sigma^2}  \right]} \right)
- \mu_n x_{mu}^{nb} - \nu_{mu}^{nb} \\  
 \leq 0,  \text{if} \  p_{mu}^{nb} \geq 0, \forall n \in \mathcal{N}, \forall b \in \mathcal{B}, \forall m \in \mathcal{M}, \forall u \in \mathcal{U}.  
\end{split} 
\end{equation} 
When $p_{mu}^{nb} >0, \frac{\partial \mathcal{L}(\boldsymbol{P}, \boldsymbol{\lambda}, \boldsymbol{\mu}, \boldsymbol{\nu})}{\partial p_{mu}^{nb}}=0$. Therefore, 
\begin{equation}
\begin{split}
    \left( \frac{\delta_{m}^ux_{mu}^{nb}g_0}{\ln(2)(d_{mu}^n)^{\alpha} \sigma^2\left[ 1+ \frac{p_{mu}^{nb}g_0}{(d_{mu}^n)^{\alpha} \sigma^2}  \right]}\right) - \theta_n - \mu_n x_{mu}^{nb} - \nu_{mu}^{nb}\\
   +\lambda_m^u\left(\frac{x_{mu}^{nb}g_0}{\ln(2)(d_{mu}^n)^{\alpha} \sigma^2\left[ 1+ \frac{p_{mu}^{nb}g_0}{(d_{mu}^n)^{\alpha} \sigma^2}  \right]} \right) = 0.
\end{split} 
\end{equation}
Finally, by doing numerical calculation, we can obtain:
\begin{equation}
    p_{mu}^{nb*}= \left[ \frac{x_{mu}^{nb}\omega(\delta_m^u  + \lambda_m^u)}{\theta_n + \mu_n x_{mu}^{nb} + \nu_{mu}^{nb}} - \frac{(d_{mu}^n)^{\alpha}\sigma^2}{g_0}\right]^{+}.  
\end{equation}

\end{IEEEproof}

\section{Simulation Results}
\label{results}
In this section, we evaluate the performance of our proposed solution approach for the joint users association and wireless resource sharing problem in the cell-free UAVs-assisted wireless networks.

\subsection{Simulation Setup}
 The network configuration considered in this study consists of 3 MNOs with 3 UAVs and 3 PSs with 20, 10, 5 users who are distributed randomly within a 400 m $\times$ 400 m area. Moreover, UAVs owned by the MNOs are assumed to be hovering at the fixed altitude of 100 m. At each UAV, the maximum available transmit power is 35 dBm, the maximum available subchannels at each UAV is 20 where each subchannel has the total bandwidth of 150 kHz, the noise density is considered as -174 dBm/Hz, and the channel gain at the reference distance is -10 dBm. Moreover, the Rician channel fading model and free-space path loss model are adopted in this work. The unit price per subchannel and transmit power set by the MNOs are between [2, 3] and [4, 5], respectively. The minimum rate requirement of each user in each SP is between [20, 30] Mbps. Finally, the payment (unit price per Mbps) of each mobile user to its associated SP is between [0.3, 0.5].

\subsection{Detailed Numerical Results}
This subsection focuses primarily on the performance improvement of our proposed algorithm. Firstly, we compare the performance of our proposed solution approach to the performance of the benchmark schemes, which are as follows:  

\begin{itemize}
	\item   RCOP: In this method, the subchannels available at each UAV are randomly allocated to the users of SPs. However, the transmit power of the UAV is allocated to the users by using our proposed Lagrangian relaxation approach.
	
	\item ECOP: In this scheme, the subchannels of each UAV are equally allocated among its associated users who belong to different SPs, and our proposed Lagrangian relaxation-based solution approach is being used to allocated the UAV's transmit power the the mobile users.

	\item RPOC: In this approach, the transmit power of each UAV is randomly allocated to its associated users, and meanwhile, the available subchannels of each UAV are allocated to its associated users by deploying our proposed whale optimization algorithm.

	\item EPOC: In this design, the transmit power of the UAV is equally allocated among its associated users who belong to different SPs. Then, the subchannels of each UAV are allocated to users by adopting our proposed whale optimization approach.  
        
\end{itemize}

Moreover, we also compare the performance of our proposed solution approach with Generalized Kelly Mechanism which was used in our previous work \cite{tun2019wireless}, and Kelly Mechanism (KM) \cite{kelly1997charging}.  

\begin{figure}[t!]
	\centering
	\captionsetup{justification = centering}
	\includegraphics[width=1.01\linewidth,height= 2.5in]{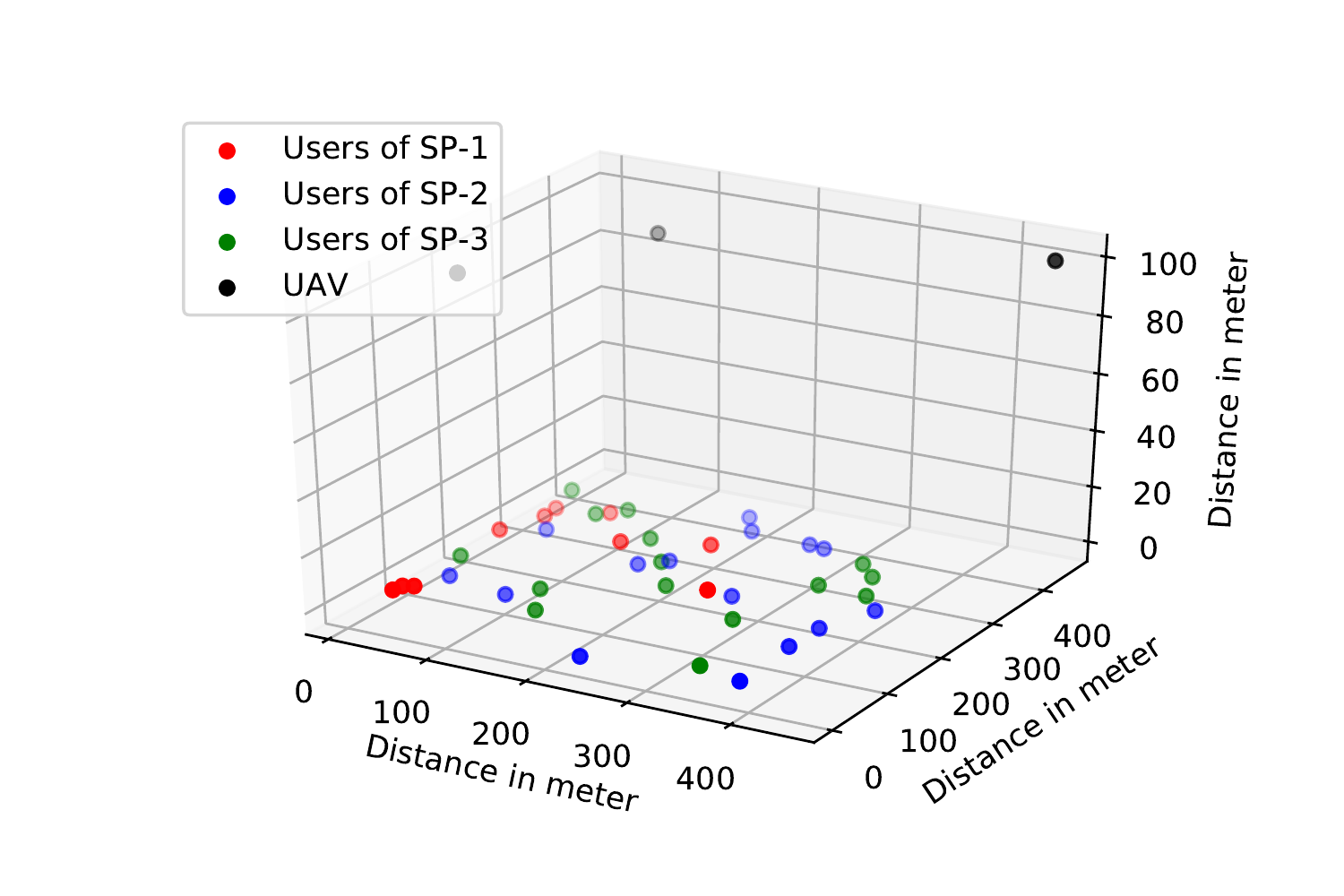}
	\caption{Network Topology.}
	\label{Network Topology.}
\end{figure}

\begin{figure}[t!]
	\centering
	\captionsetup{justification = centering}
	\includegraphics[width=\linewidth]{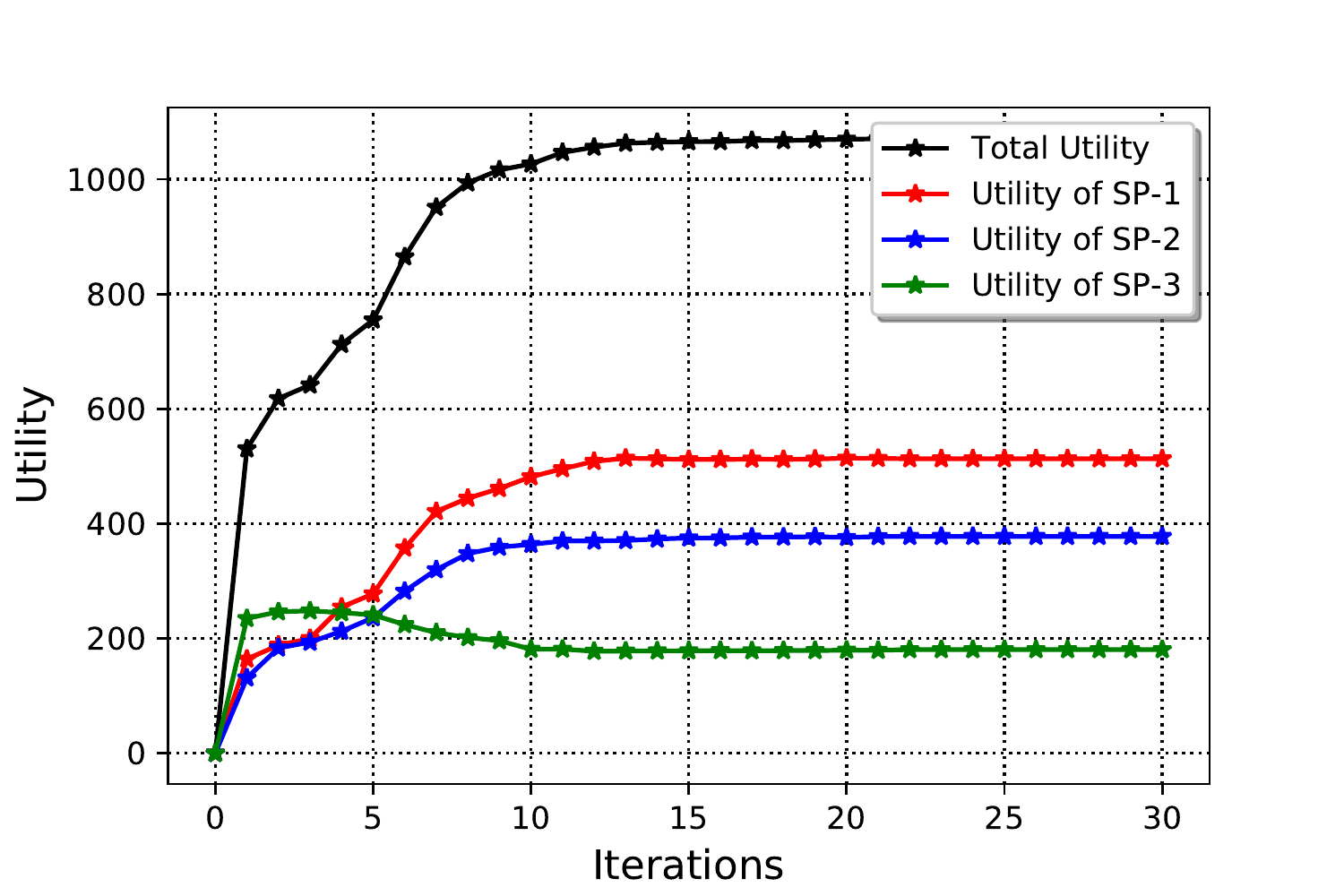}
	\caption{Convergence of utility of SPs under the proposed algorithm.}
	\label{Convergence}
\end{figure}

Fig.~\ref{Network Topology.} demonstrates the network topology of our work which consists of 3 UAVs and 35 mobile users. Fig.~\ref{Convergence} depicts the convergence of the utility of each SP and the total utility in the network. From the figure, we observe that our proposed algorithm converges to the solution in lesser than 22 iterations. Therefore, our proposed algorithm is suitable to implement in real network environment. Furthermore, when compared to other SPs, we find that SP-1 achieves the highest utility. The explanation for this is that SP-1 has more users than SP-2 and SP-3, and that the payment for each user established by SP-1 is higher than the other SPs.

\begin{figure}[h!]
	\centering
	\captionsetup{justification = centering}
	\includegraphics[width=3.8in, height=2.7in]{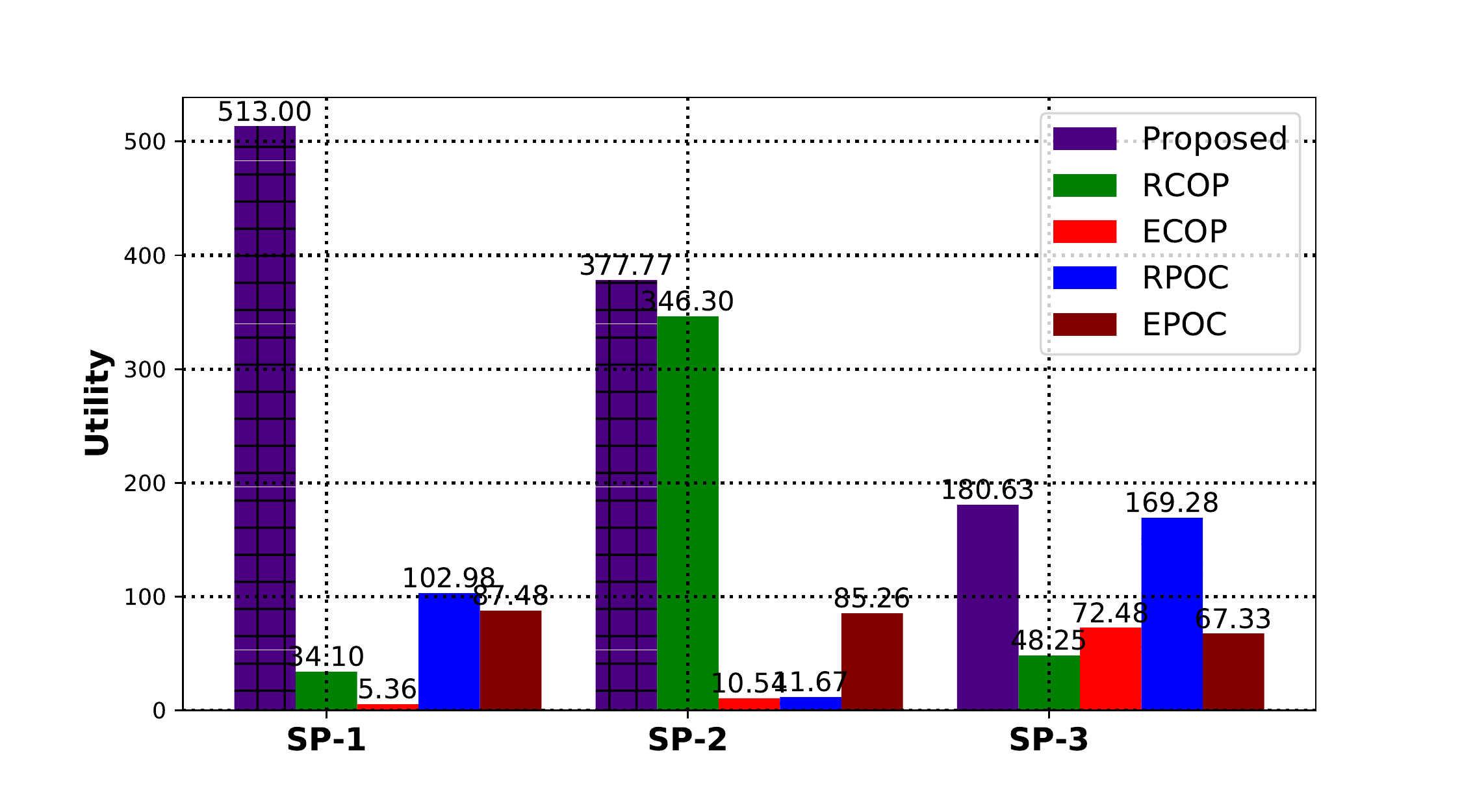}
	\caption{Comparison of achieved utility of SPs.}
	\label{Comparison}
\end{figure}

Fig.~\ref{Comparison} compares the achieved utility of each SP under different algorithms: proposed algorithm, RCOP method, ECOP approach, RPOC and EPOC algorithms. From Fig.~\ref{Comparison}, we observe that the utility achieved by the SP-1: 513 (Proposed), 34.1 (RCOP), 5.36 (ECOP), 102.98 (RPOC), 87.48 (EPOC), the utility of SP-2: 377.77 (Proposed), 346.3 (RCOP), 10.54 (ECOP), 11.67 (RPOC), 85.26 (EPOC), and achieved utility of SP-3: 180.63 (Proposed), 48.25 (RCOP), 72.48 (ECOP), 169.28 (RPOC), 67.33 (EPOC). From the above results, our proposed method provides a higher utility than other benchmark schemes, for all SPs. Therefore, it is obvious that our solution technique is superior than benchmark schemes.
\begin{figure}[t!]
	\centering
	\captionsetup{justification = centering}
	\includegraphics[width=1.1\linewidth]{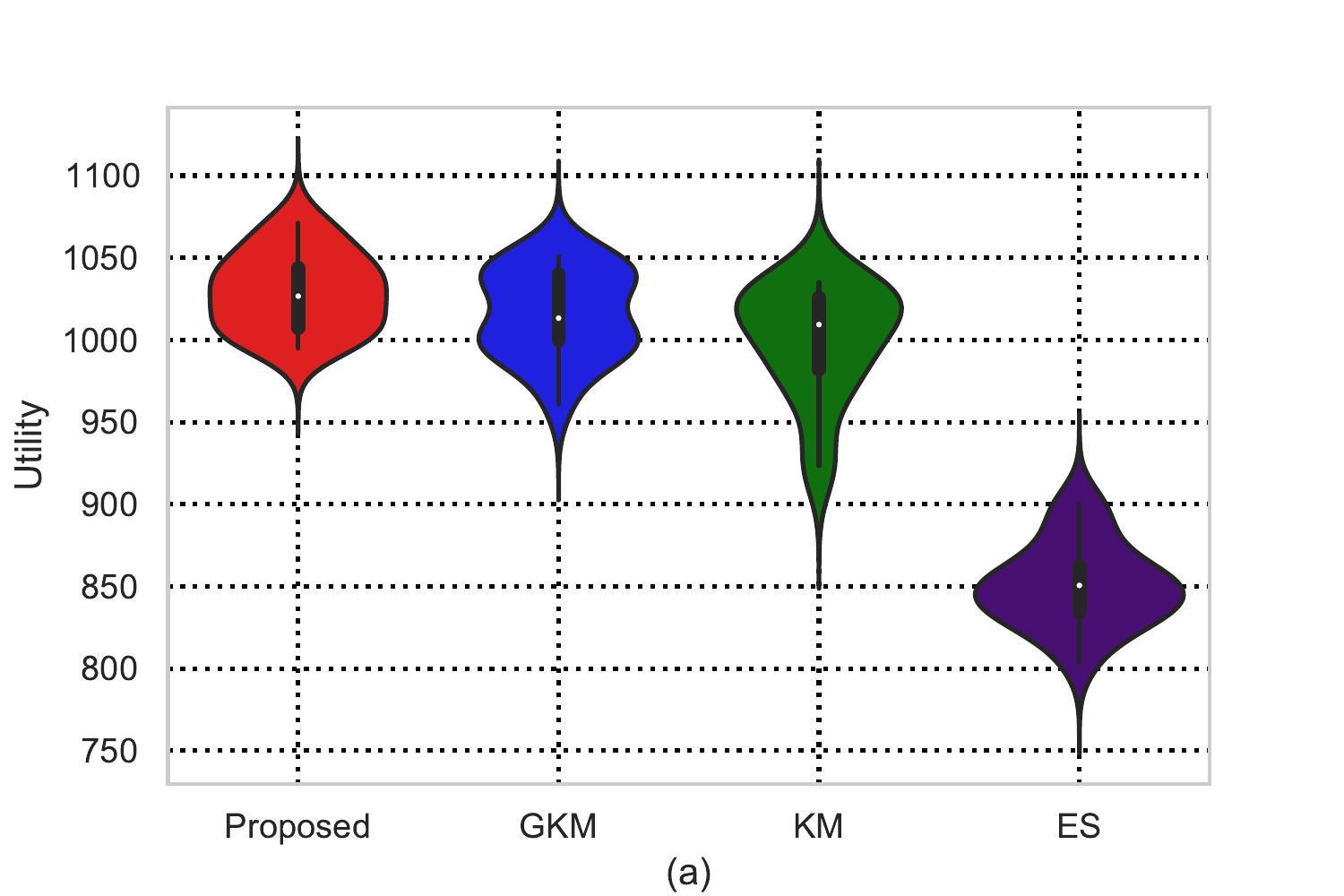}
	\caption{Comparison of achieved utility of SPs.}
	\label{Com}
\end{figure}
Moreover, in Fig.~\ref{Com}, we compare the performance of the proposed algorithm with GKM, KM, and ES schemes. From Fig.~\ref{Com}, the median of the total utility in the considered network is around 1030.76 (Proposed), 1010.32 (GKM), 1009.78 (KM), and 950 (Equal Sharing). Furthermore, the lowest and highest utility of the network is 949.96-1120 (Proposed), 900-1102 (GKM), 850-112.2 (KM), and 750-950 (ES). Thus, it is clear that the performance of our proposed algorithm outperforms existing approaches. 
 \begin{figure}[h]
	\centering
	\captionsetup{justification = centering}
	\includegraphics[width=\linewidth]{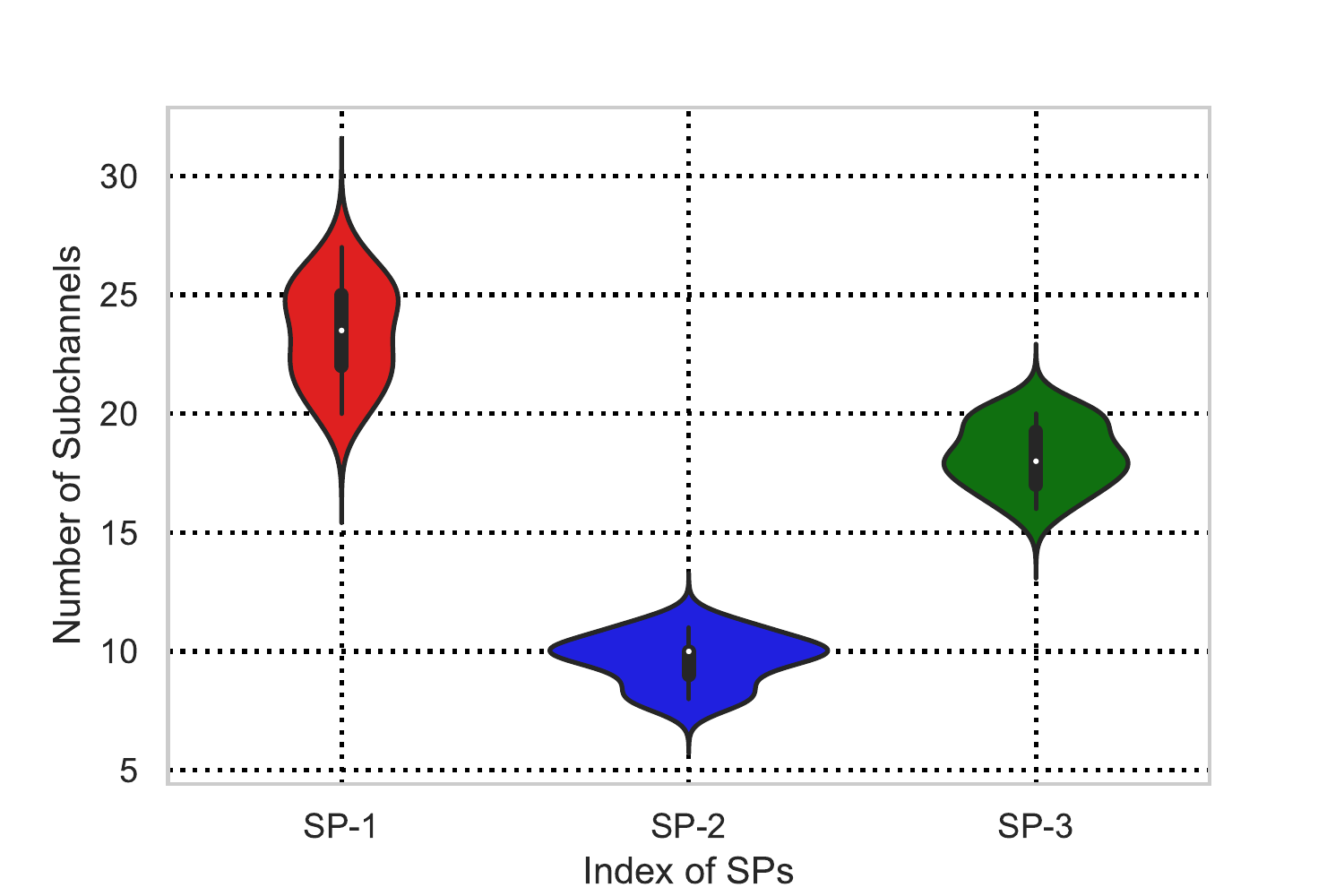}
	\caption{Number of channels allocated to each SP.}
	\label{Channel}
\end{figure}

 In Fig.~\ref{Channel}, we show the number of subchannels allocated to each SP under our proposed whale optimization algorithm by using violin plot. From Fig.~\ref{Channel}, we examine that SP-1 receives the highest number of subchannels when compares to SP-2 and SP-3. The reason is that SP-1 possesses highest number of users. Despite the fact that SP-2 has a higher number of users, the number of subchannels received by SP-2 from all UAVs is smaller than SP-3. This is due to the fact that the minimum rate requirement for SP-3 users is higher than the rate requirement for SP-2 users.

\begin{figure}[t!]
	\centering
	\captionsetup{justification = centering}
	\includegraphics[width=\linewidth]{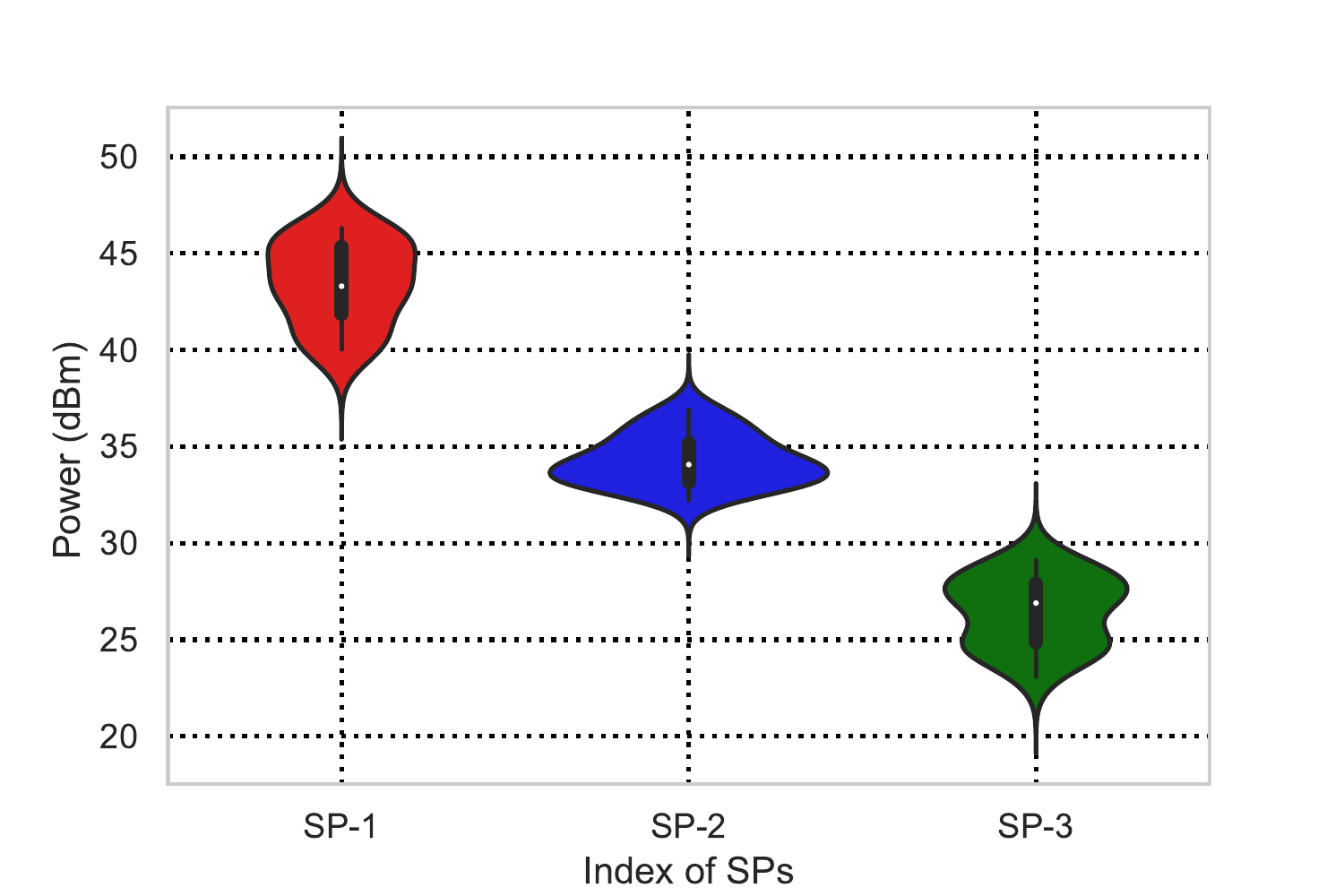}
	\caption{Power allocated to each SP.}
	\label{power}
\end{figure}

\begin{figure}[H]
	%\vspace{-0.9cm}
	\centering
	\begin{subfigure}{0.3\textwidth} % width of left subfigure
		\includegraphics[width=\linewidth]{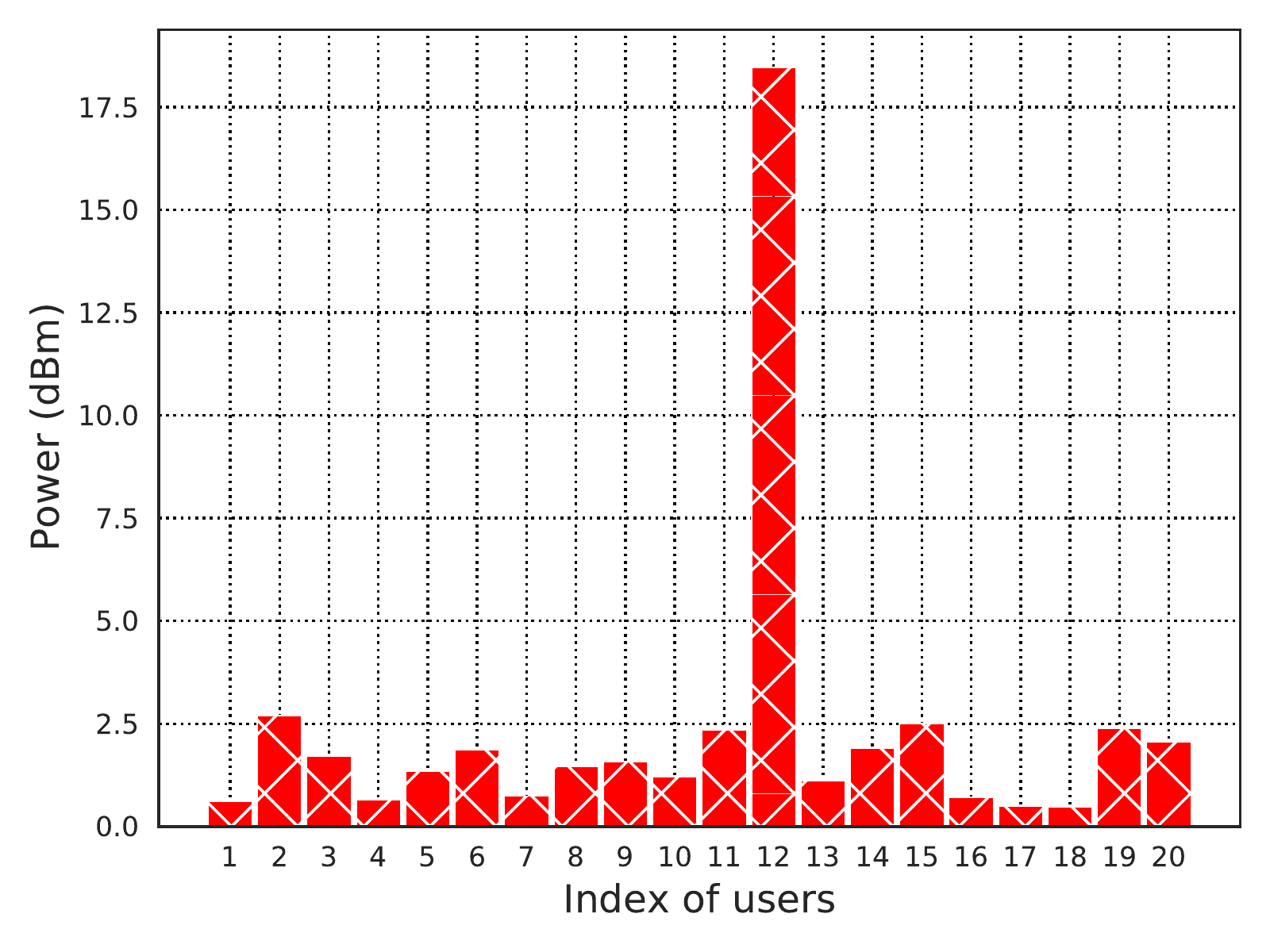}
		\caption{Power allocation to each user of SP-1.}
		\label{12}
	\end{subfigure}\hfil
	%	\hspace{1em} % here you can insert horizontal or vertical space
	\begin{subfigure}{0.3\textwidth} % width of right subfigure
		\includegraphics[width=\linewidth]{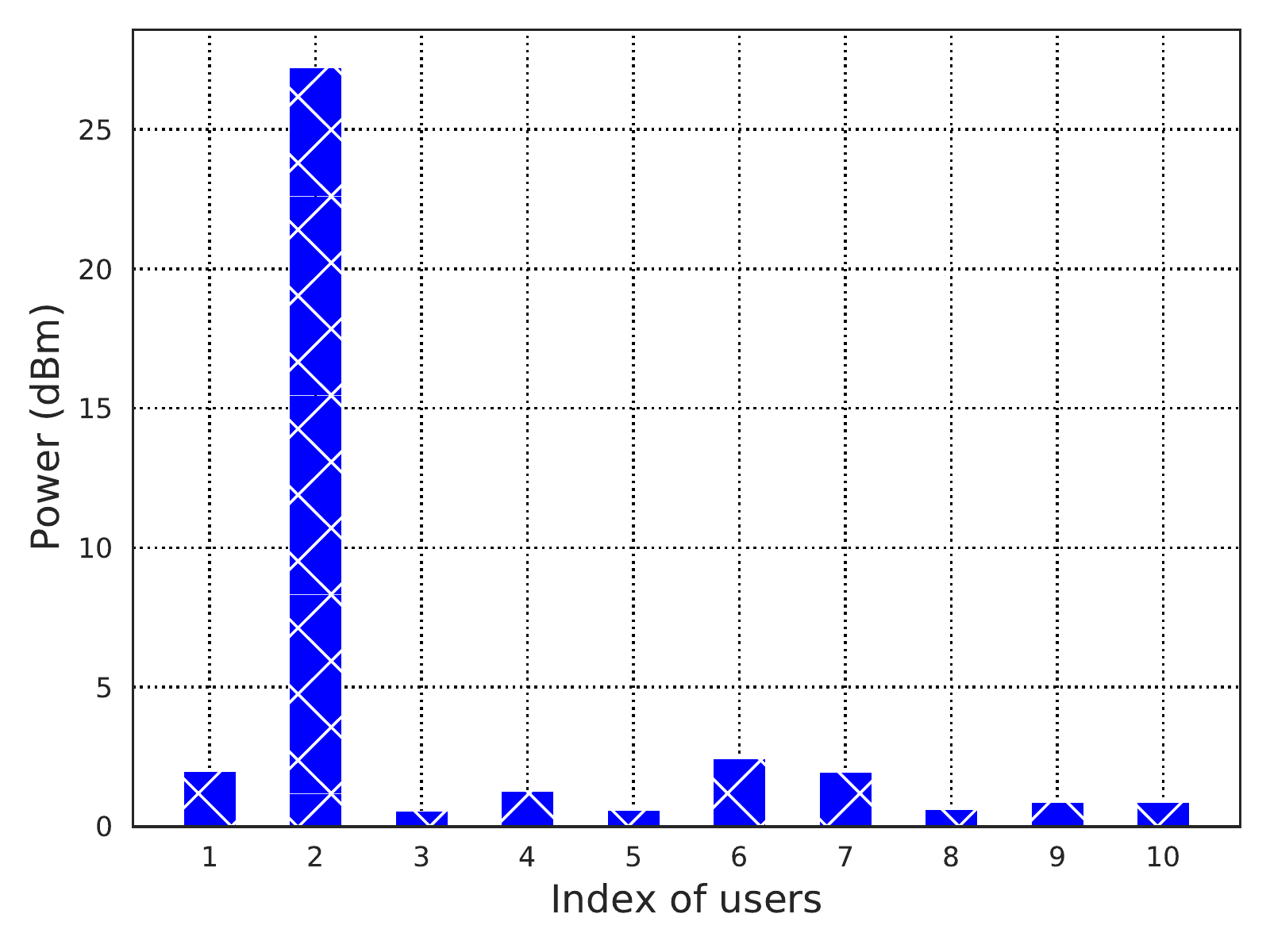}
		\caption{Power allocation to each user of SP-2.}
		\label{13}
	\end{subfigure} \hfil
	%	\hspace{1em} % here you can insert horizontal or vertical space
	\begin{subfigure}{0.3\textwidth} % width of right subfigure
		\includegraphics[width=\linewidth]{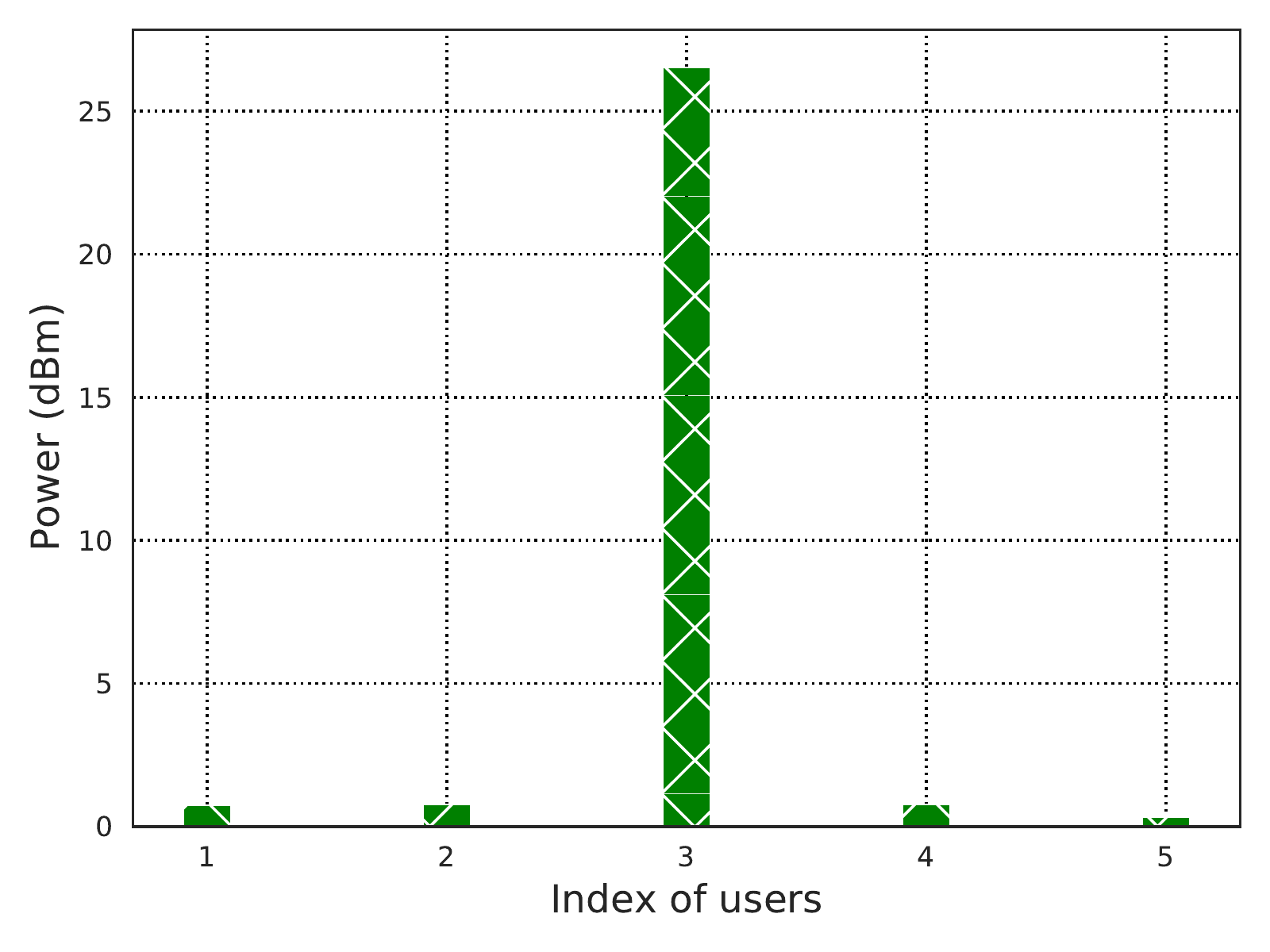}
		\caption{Power allocation to each user of SP-3.}
		\label{14}
	\end{subfigure}
	%\vspace{-0.2cm}
	\caption {Power allocation to users of SPs.} \label{ffff}
\end{figure}

\begin{figure}[t!]
	\centering
	\captionsetup{justification = centering}
	\includegraphics[width=\linewidth]{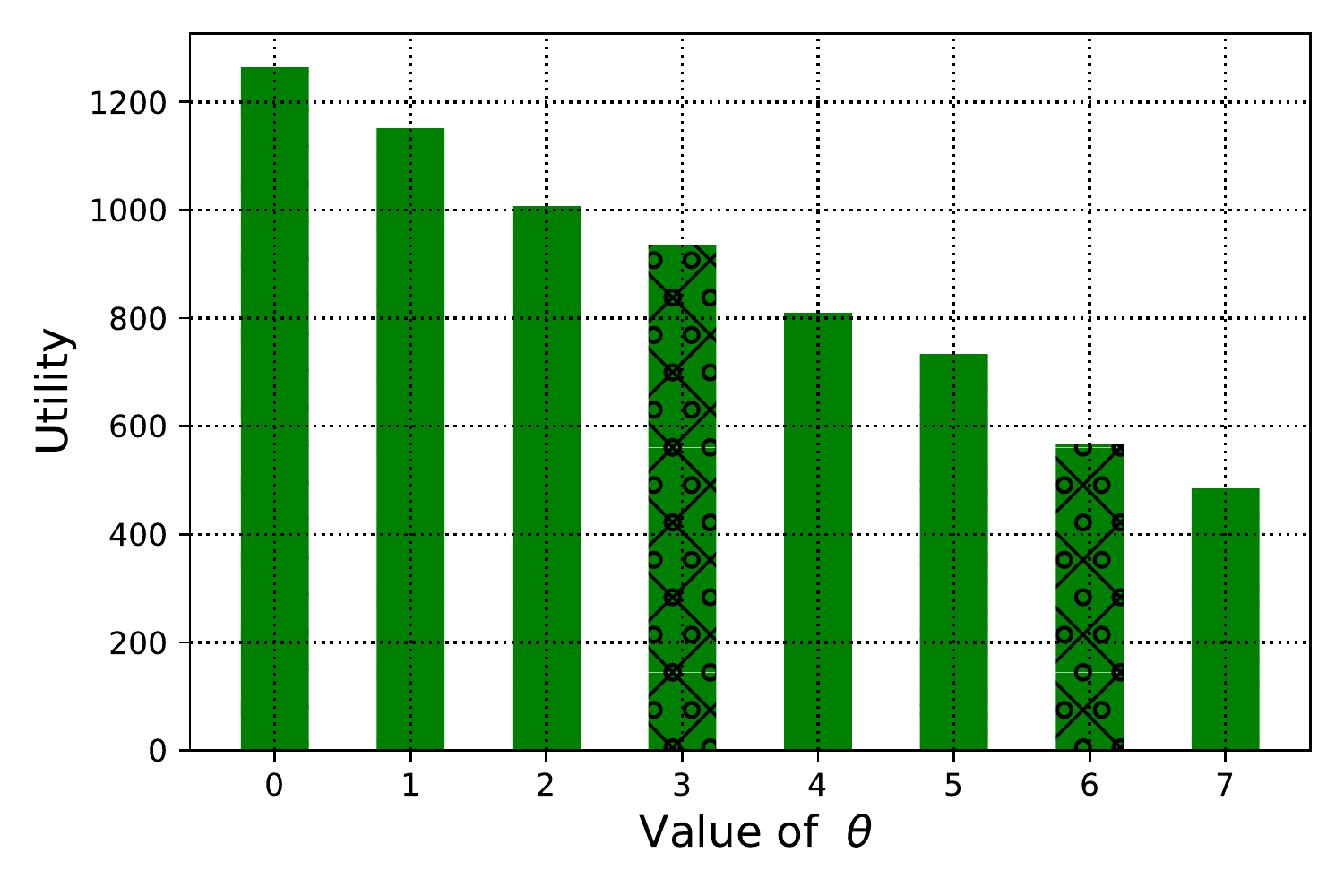}
	\caption{Utility versus value of $\boldsymbol{\theta}$.}
	\label{theta}
\end{figure}

\begin{figure}[t]
	\centering
	\captionsetup{justification = centering}
	\includegraphics[width=\linewidth]{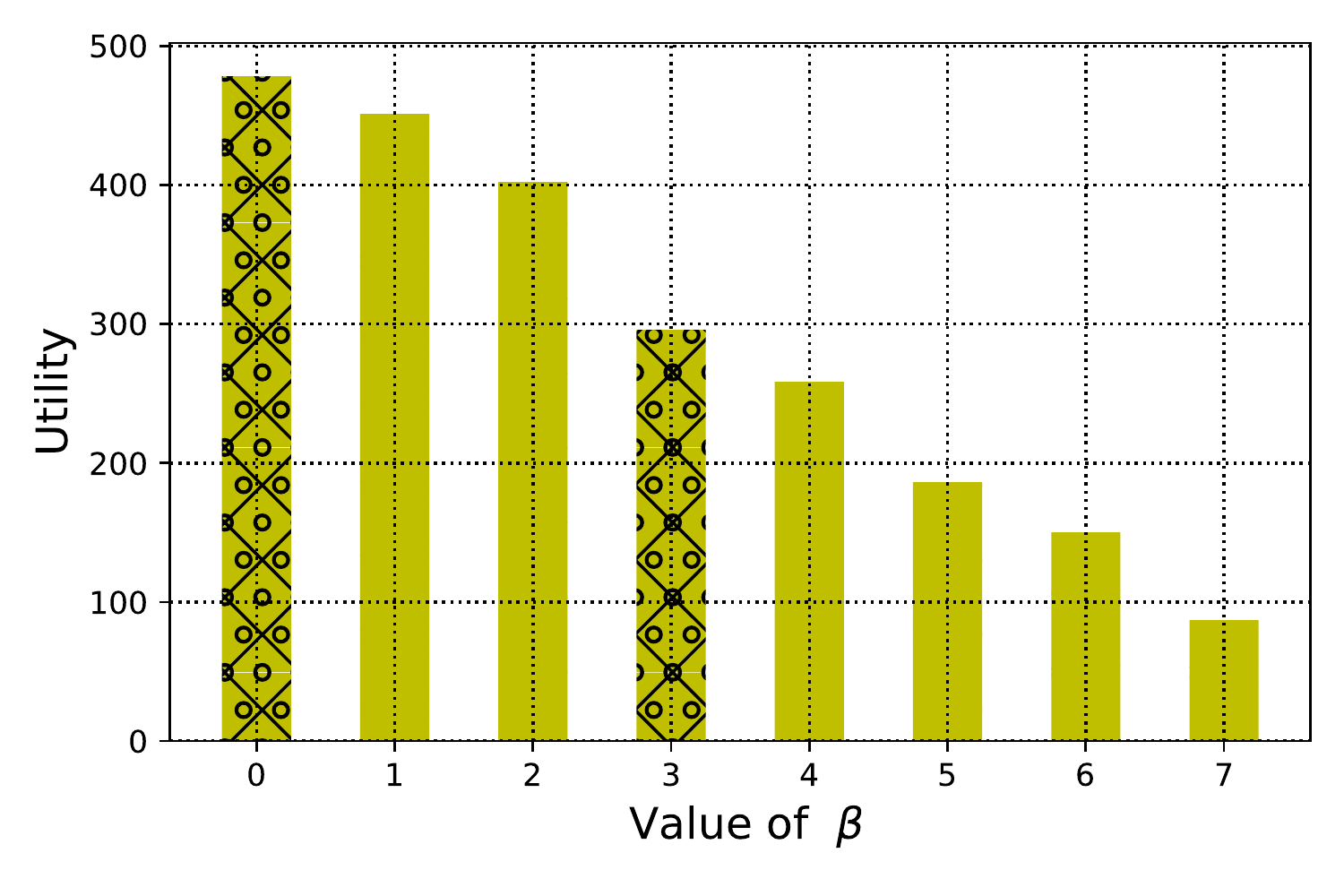}
	\caption{Utility versus value of $\boldsymbol{\beta}$.}
	\label{beta}
\end{figure}
 
 \begin{figure}[t]
 	\centering
 	\captionsetup{justification = centering}
 	\includegraphics[width=\linewidth]{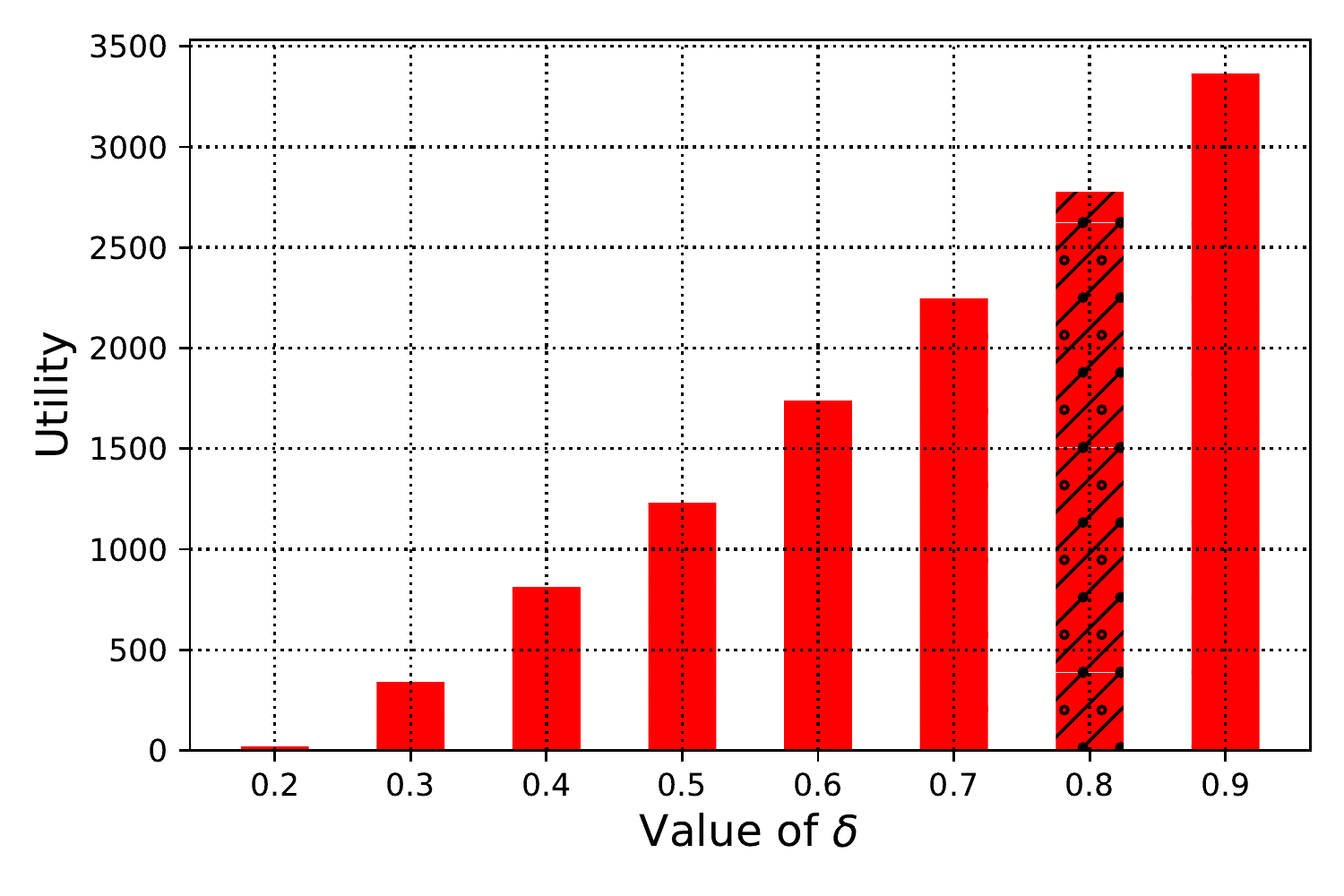}
 	\caption{Utility versus value of $\boldsymbol{\delta}$.}
 	\label{delta}
 \end{figure}
 \begin{figure}[t]
 	\centering
 	\captionsetup{justification = centering}
 	\includegraphics[width=\linewidth]{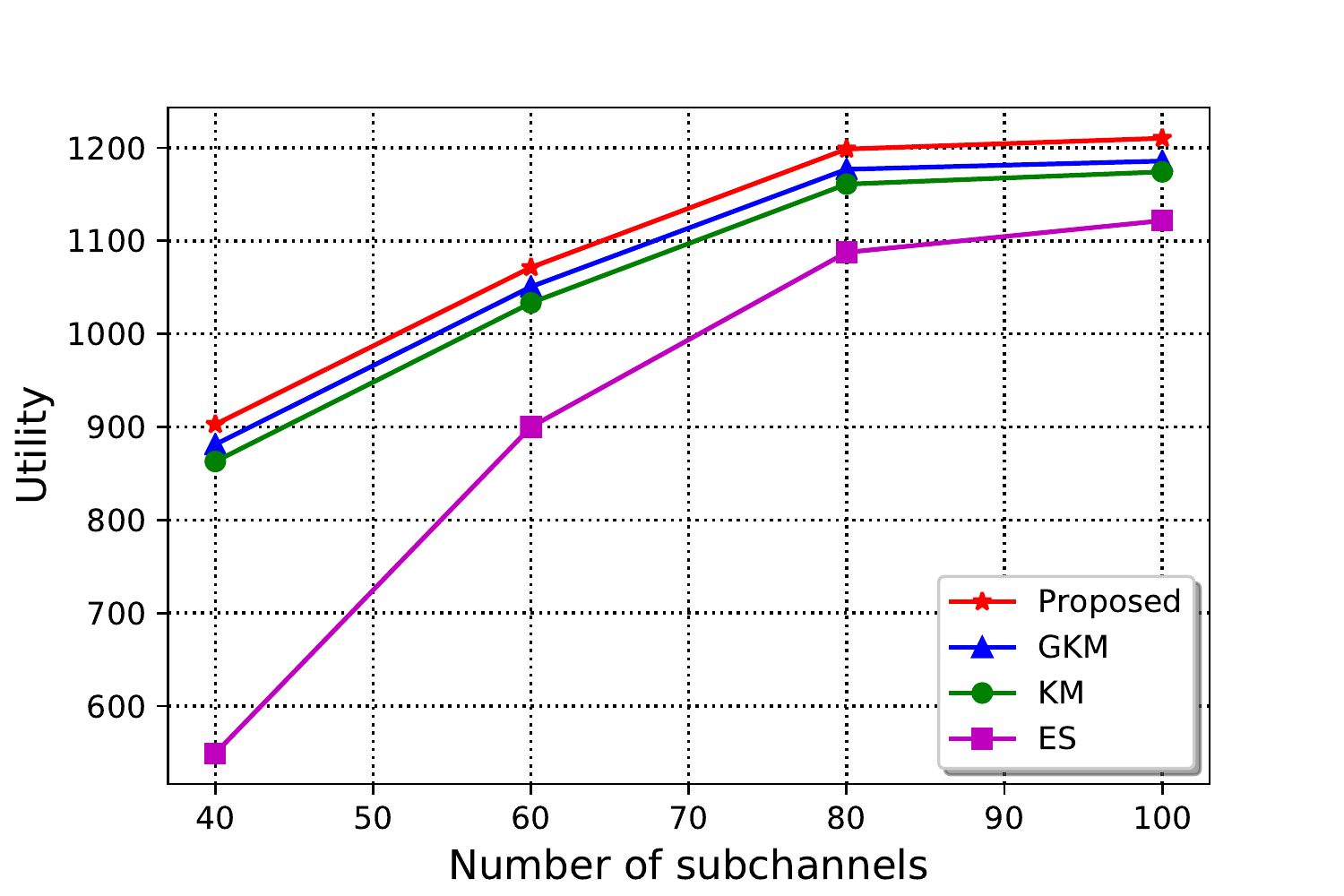}
 	\caption{Utility versus number of subchannels.}
 	\label{CP}
 \end{figure}
 
 Moreover, Fig.~\ref{power} demonstrates that the transmit power of the UAVs allocated to each SP under our proposed algorithm. From the figure, we observe that the median of the transmit power of the UAVs gained by SP-1 is 43.6 dBm, SP-2 is 33.86 dBm, and SP-3 is 27.12 dBm. Furthermore, from Fig.~\ref{power}, we can examine that the highest and lowest transmit power received by SP-1 is 51.94-35.03 dBm, SP-2 is 39.95-29.57 dBm, and SP-3 is 33.85-19.53 dBm. Therefore, SP-1 achieves higher portion of the transmit power of the UAVs. It is because SP-1 has more users and the minimum rate requirement of users in SP-1 is higher than rate requirement of users in SP-2 and SP-3. Then, Fig.~\ref{ffff} represents the power allocation to each user of each SP. From Fig.~\ref{12}, we can observe that amongst the users of SP-1, user-12 and user-17 receive the largest and smallest portion of the transmit power of their associated UAVs, respectively. It is because the achievable channel gain of the user-12 and user-17 are the weakest and strongest amongst users of SP-1. Moreover, it is also possible that the minimum rate requirement of user-12 and user-17 are the highest and the lowest. We can easily see that above mentioned reasons in problem \textbf{P22}. Similarly, Fig.~\ref{13} and Fig.~\ref{14} demonstrate the power allocation to each user of SP-2 and SP-3 where we observe that user-2 in SP-2 and user-3 in SP receive the largest portion of the UAVs' transmit power.     

Fig.~\ref{theta} demonstrates the effect of the unit price per subchannel set by MNOs on the the total network utility. we observe from  Fig.~\ref{theta} that as the value of $\boldsymbol{\theta}$ decreases, so does the total network utility. The reason for this is that as MNOs increase the unit per subchannel, SPs need to pay more to MNOs, resulting in an increase in SPs' costs. As a consequence, the utility of all SPs diminishes (i.e., the total network utility decreases). Moreover, we also present the effect of the unit price per transmit power on the total unit utility in Fig.~\ref{beta}. From the figure, we observe that the value of the total network utility decreases when we increase the value of $\boldsymbol{\beta}$. Finally, Fig.~\ref{delta} depicts the impact of the value of $\boldsymbol{\delta}$ on the total network utility. The total network utility increases as SPs increase the payment $\boldsymbol{\delta}$ (unit price per Mbps) for their mobile users, as seen in Fig.~\ref{delta}. It goes without saying that as users pay more to SPs, the revenue of SPs $U_m^{\textrm{Rev}}, \forall m \in \mathcal{M}$ rises. As a result, the utility of the SPs will increase. In other words, the total network utility increases.

Finally, in Fig.~\ref{CP}, we show the total network utility for various numbers of available subchannels in the network. We can see from Fig.~\ref{CP} that the network utility rises as the total number of subchannels rises. Furthermore, our proposed algorithm outperforms GKM, KM, and ES methods, as shown in the figure.

\section{Conclusion}
\label{con} 

In this paper, we have proposed cell-free UAVs-assisted wireless networks in which SPs share the wireless resources of the MNOs. Then, we have formulated the joint users association and resource sharing problem of the proposed model with the objective of maximizing the total network utility of SPs. Since the formulated problem is a mixed-integer, nonlinear, and non-convex problem, to be tractable, the formulated problem was decomposed into two subproblems: users association and resource sharing problem. Then, we deployed a two-sided matching algorithm in order to solve the users association problem. Moreover, we also applied the whale optimization algorithm and Lagrangian relaxation method to solve the resource sharing problem. Finally, we have presented extensive numerical results to validate the efficacy of our proposed solution approach that outperforms the other benchmark schemes and existing algorithms. For future direction, we will integrate reconfigurable intelligent surfaces (IRSs) technology in the UAVs-assisted wireless network in order to improve the system throughput.

% or


\ifCLASSOPTIONcaptionsoff
  \newpage
\fi



%\bibliographystyle{IEEEtran}
\bibliographystyle{IEEEtran}
\bibliography{MNO}



%\vfill


\end{document}